\documentclass{IEEEtran}

\usepackage{amsmath}
\usepackage{amsfonts}
\usepackage{siunitx}
\usepackage{graphicx}
\usepackage[nocompress]{cite}
\usepackage{booktabs}
\usepackage{microtype}
\usepackage{url}
\usepackage[caption=false,font=footnotesize]{subfig}
\usepackage{hyperref}
\usepackage[export]{adjustbox}
\usepackage{orcidlink}

\DeclareSIUnit{\nothing}{\relax}

\DeclareMathOperator{\sgn}{sgn}

\makeatletter
\def\bstctlcite{\@ifnextchar[{\@bstctlcite}{\@bstctlcite[@auxout]}}
\def\@bstctlcite[#1]#2{\@bsphack
  \@for\@citeb:=#2\do{%
    \edef\@citeb{\expandafter\@firstofone\@citeb}%
    \if@filesw\immediate\write\csname #1\endcsname{\string\citation{\@citeb}}\fi}%
  \@esphack}
\makeatother

\begin{document}

\title{End-to-End Multi-Task Learning for Adjustable Joint Noise Reduction and Hearing Loss Compensation}

\author{Philippe~Gonzalez\,\orcidlink{0009-0006-4965-3514}, Vera~Margrethe~Frederiksen\,\orcidlink{0009-0007-6796-6175}, Torsten~Dau\,\orcidlink{0000-0001-8110-4343}, Tobias~May\,\orcidlink{0000-0002-5463-5509}\thanks{The authors are with the Department of Health Technology, Technical University of Denmark, 2800 Lyngby, Denmark (e-mail: phigon@dtu.dk; vmbfr@dtu.dk; tdau@dtu.dk; tobmay@dtu.dk).}}

\robustify\bfseries

\markboth{Submitted to IEEE Transactions on Audio, Speech, and Language Processing}{}

\maketitle

\begin{abstract}
A multi-task learning framework is proposed for optimizing a single deep neural network (DNN) for joint noise reduction (NR) and hearing loss compensation (HLC).
A distinct training objective is defined for each task, and the DNN predicts two time-frequency masks.
During inference, NR and HLC can be adjusted independently by exponentiating each mask before combining them.
In contrast to recent approaches that rely on training an auditory-model emulator to define a differentiable training objective, we propose an auditory model that is inherently differentiable, thus allowing end-to-end optimization.
The audiogram is provided as an input to the DNN, thereby enabling listener-specific personalization without the need for retraining.
Results show that the proposed approach not only allows adjusting NR and HLC independently, but also improves objective metrics compared with optimizing a single training objective.
It also outperforms a cascade of two DNNs separately trained for NR and HLC and shows competitive HLC performance against a traditional hearing-aid prescription on objective metrics.
To the best of our knowledge, this is the first study that uses an auditory model to train a single DNN for both NR and HLC across a wide range of flat and high-frequency-sloping hearing-loss profiles.
\end{abstract}

\begin{IEEEkeywords}
Differentiable auditory model, multi-task learning, noise reduction, hearing loss compensation
\end{IEEEkeywords}

\section{Introduction}
\label{sec:introduction}

\IEEEPARstart{H}{earing} impairment affects millions of people worldwide and has major negative consequences for social interaction, cognitive function, and overall quality of life~\cite{world2021world}.
Mild-to-moderate sensorineural hearing loss is typically treated with hearing aids, which aim to improve audibility and listening comfort through noise reduction (NR) and hearing loss compensation (HLC) algorithms.
Although these algorithms can be effective, hearing-impaired (HI) listeners often continue to report difficulties understanding speech in complex acoustic environments, even when wearing hearing aids.
Moreover, the typical signal processing building blocks used in hearing aids, such as beamforming, NR, and amplification, are often designed and optimized independently~\cite{georganti2020intelligent}.
In contrast, deep neural networks (DNNs) can jointly model multiple signal processing stages and may surpass traditional approaches by learning complex nonlinear relationships between noisy speech, listener profiles, and optimal NR and HLC strategies.

In recent years, DNNs have been successfully applied to speech enhancement and NR~\cite{wang2018supervised,zheng2023sixty}.
These systems are typically trained by generating noisy speech mixtures from clean speech utterances and noise segments.
As a result, the clean speech ground-truth target is directly available during training, making it straightforward to define a training objective.
In contrast, DNN-based HLC remains largely unexplored and is considerably more challenging due to the absence of an obvious ground-truth target.
The recent Clarity Enhancement Challenge~\cite{graetzer2021clarity} addressed this issue by requiring submitted systems to perform joint NR and HLC.
The systems were then evaluated subjectively by a panel of HI listeners.
Despite this effort, most submissions focused primarily on developing strong NR systems and combined them with generic HLC strategies such as NAL-R and dynamic range compression.
This underlines the current lack of DNN-based HLC solutions that can outperform conventional hearing-aid prescriptions.

One way to train a DNN for HLC is to minimize the difference between the outputs of a normal-hearing (NH) auditory model and an HI auditory model.
More specifically, the NH auditory model provides a physiologically grounded computational target, and the DNN is tasked with processing the input signal such that the output of the HI auditory model matches this target.
However, off-the-shelf auditory models are not well suited to this application, since they are nondifferentiable and thus prevent gradients from being backpropagated to update the DNN parameters.
Additionally, they operate at very high sampling rates and involve multiple processing stages, which makes them computationally expensive.
To address these limitations, recent studies have trained auxiliary DNNs to approximate nondifferentiable models of the auditory periphery~\cite{baby2021convolutional,drakopoulos2021convolutional,nagathil2021computationally,nagathil2023wavenet,leer2024train}.
These auditory-model emulators were then used as proxies for defining differentiable training objectives~\cite{drakopoulos2022differentiable,drakopoulos2023neural,drakopoulos2023dnn,wen2025dconnear,wen2025individualized,leer2025hearing}.
However, this approach requires training an additional DNN.
As a result, it becomes difficult to investigate the effect of individual auditory-model parameters on HLC performance, since any update to the auditory model requires retraining the emulator.
Moreover, most studies have trained both the auditory-model emulator and the HLC system for individual listener profiles.
This limits practical applicability, since training two DNNs for each hearing-aid user is time-consuming.

An alternative approach is to design auditory models that are inherently differentiable.
Such models can be integrated directly into the optimization framework of the HLC algorithm, thus enabling end-to-end training.
This makes it easier to investigate which auditory-model stages are essential for delivering perceptual benefits, since no additional emulator needs to be retrained whenever the auditory model is modified.
Previous work has shown that HLC algorithms trained in this way~\cite{tu2021dhasp,tu2021optimising,tu2021two} can yield improved objective metrics compared with traditional hearing-aid prescriptions, both in quiet~\cite{tu2021dhasp} and in noise~\cite{tu2021optimising}.
However, these studies trained systems for individual listener profiles, and it therefore remains unclear whether the results generalize to systems trained for a broad range of HI listeners.
In addition, they used relatively simple learning-based HLC algorithms with few trainable parameters, such as fixed banks of finite impulse response (FIR) filters with learnable gains.
To provide benefits across a wide range of HI listeners, deeper neural networks may be required, yet such models remain largely unexplored for end-to-end HLC learning.

A small number of studies have proposed solutions that can be personalized without training the HLC system, or an auditory-model emulator, separately for each listener profile~\cite{drgas2024dynamic,zmolikova2021but,cheng2023speech,ni2026affine}.
This is typically achieved by providing the user's audiogram as an additional input to the DNN, although this does not allow compensation for supra-threshold deficits.
In~\cite{drgas2024dynamic}, different HLC algorithms were trained using a differentiable implementation of the Moore, Stone, Baer, and Glasberg (MSBG) model~\cite{baer1993effects,baer1994effects,moore1993simulation,stone1999tolerable} proposed in~\cite{tu2021optimising}.
However, joint NR and HLC was not considered.
In~\cite{zmolikova2021but}, the same differentiable MSBG model was used, but separate NR and HLC stages were optimized and applied sequentially.
This mirrors the typical hearing-aid processing chain, in which NR is applied before HLC.
However, optimizing each task in isolation may lead to suboptimal performance or unnecessary redundancies.
Since NR and HLC are both audio-to-audio tasks, internal representations learned by the DNN for one task may also benefit the other.
Jointly optimizing a single DNN for both tasks may therefore be more efficient and may produce more robust internal representations~\cite{baxter2000model}.
In~\cite{cheng2023speech,ni2026affine}, a single DNN was trained to perform both NR and HLC.
However, the training target was clean speech processed by a traditional hearing-aid prescription, namely the FIG6 formula~\cite{killion1993types}.
This approach implicitly assumes that FIG6 is the optimal HLC strategy and trains the DNN to approximate it.
In contrast, training the system to restore the output of a physiologically grounded auditory model may allow the discovery of HLC strategies that outperform traditional prescriptions.

Another limitation of previous single-DNN solutions for joint NR and HLC is that they do not allow adjusting NR or HLC at inference time.
Studies have shown that some subpopulations of HI listeners prefer strong NR, whereas others prefer milder NR~\cite{neher2014relating,neher2016directional}.
More generally, users may wish to prioritize NR or HLC depending on the acoustic environment or their listening intent.
For example, in very noisy environments, users may prefer strong NR, whereas in small social gatherings, HLC without NR may be more desirable.
The ability to adjust NR and HLC independently at inference time is therefore a desirable feature in hearing aids.
Previous single-DNN approaches~\cite{tu2021optimising,leer2025hearing,cheng2023speech,ni2026affine} used training objectives whose individual terms optimized both tasks simultaneously.
As a consequence, it is unclear which task drives the optimization, or whether the DNN implicitly prioritizes one task over the other.
One recent study addressed this issue by providing the target signal-to-noise ratio (SNR) as an additional input to the DNN~\cite{wen2025individualized}.
However, that system could not adjust HLC and was still trained for individual listener profiles.

In the present work, we propose a DNN that performs joint NR and HLC for a wide range of flat and high-frequency-sloping hearing-loss profiles.
This is achieved through a multi-task learning framework and a differentiable auditory model, which together eliminate the need for an auditory-model emulator and enable end-to-end optimization.
Each task is assigned a distinct training objective, and the two objectives are automatically balanced during training using an uncertainty-based weighting scheme~\cite{kendall2018multi}.
During inference, the time-frequency (TF) masks predicted for each task are combined after being exponentiated by two independent parameters, thereby enabling independent adjustment of NR and HLC.
This represents a significant improvement over our previous study~\cite{gonzalez2025controllable}, in which the denoised and compensated signals were mixed in the time domain using a single parameter.
The user's audiogram is provided as an input to the DNN, allowing personalization without retraining the system for each listener.
Different objective metrics are used to compare the proposed approach with both a traditional hearing-aid prescription and a cascade of two DNNs optimized separately for each task.
Code and audio examples are available online\footnote{\url{https://philgzl.com/cnrhlc}}.

\section{Previous frameworks}
\label{sec:previous}

\noindent This section describes previous approaches for training a DNN for NR-only, HLC-only, or joint NR and HLC using auditory models.
We consider the case where a single DNN is optimized, and therefore exclude approaches where separate NR and HLC systems are trained and applied sequentially~\cite{tu2021two,zmolikova2021but}.

Let~$\mathcal{F}_\theta$ be a DNN that takes a signal~$x$ as input and is optimized over its parameters~$\theta$ to perform NR, HLC, or joint NR and HLC.
Let~$\mathcal{A}_\mathrm{NH}$ be an NH auditory model and~$\mathcal{A}_\mathrm{HI}$ an HI auditory model.
These can be auditory models that are differentiable by design, or emulators that were previously trained to approximate nondifferentiable auditory models.
$\mathcal{A}_\mathrm{HI}$ is parameterized by an audiogram~$\Delta^\mathrm{dB}$ that characterizes the hearing loss of a given HI listener.
The audiogram~$\Delta^\mathrm{dB}$ can be provided as an additional input to~$\mathcal{F}_\theta$ to enable personalization without the need for retraining for each listener profile.

\subsection{NR-only}
\label{sec:previous_nr}

\begin{figure}
  \centering
  \subfloat[NR-only]{
    \includegraphics[scale=0.97]{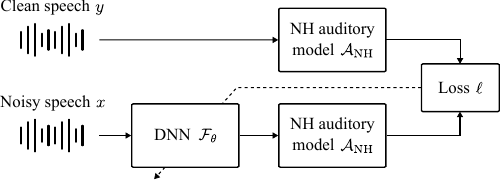}
    \label{fig:previous_nr}
  }\\
  \subfloat[HLC-only]{
    \includegraphics[scale=0.97]{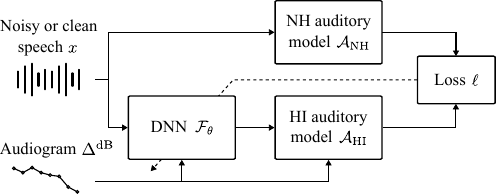}
    \label{fig:previous_hlc}
  }\\
  \subfloat[Joint NR and HLC]{
    \includegraphics[scale=0.97]{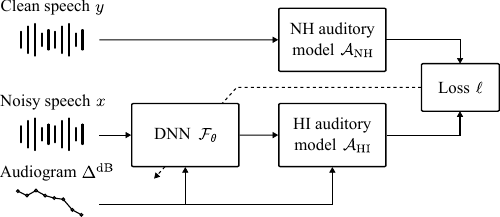}
    \label{fig:previous_nr_hlc}
  }
  \caption{Traditional single-DNN frameworks for (a) NR-only, (b) HLC-only, and (c) joint NR and HLC using auditory models.}
  \label{fig:previous}
\end{figure}

\noindent Figure~\ref{fig:previous_nr} describes how~$\mathcal{A}_\mathrm{NH}$ can be used to train~$\mathcal{F}_\theta$ for NR-only.
The denoised output signal~$\hat{y}=\mathcal{F}_\theta(x)$ and the corresponding clean target signal~$y$ are independently processed by~$\mathcal{A}_\mathrm{NH}$.
A loss function~$\ell$ is then used to compare the two auditory-model outputs and optimize~$\mathcal{F}_\theta$.
The overall training objective can be expressed as
\begin{equation}
  \label{eq:loss_nr}
  \mathcal{L}_\mathrm{NR} = \ell \Bigl( \mathcal{A}_\mathrm{NH} \bigl( \mathcal{F}_\theta(x) \bigr), \mathcal{A}_\mathrm{NH}(y) \Bigr).
\end{equation}
This is conceptually similar to traditional speech enhancement, even though the vast majority of studies do not characterize their training objectives in terms of auditory modeling.
Nevertheless, commonly used training objectives can be seen as simple implementations of~$\mathcal{A}_\mathrm{NH}$ and~$\ell$, and thus technically fall into this category.
A handful of speech enhancement studies have proposed perceptually motivated training objectives that are conceptually related to auditory modeling~\cite{zhao2018perceptually,vuong2021modulation,eng2022using,monir2025frequency}.

\subsection{HLC-only}
\label{sec:previous_hlc}

\noindent Figure~\ref{fig:previous_hlc} describes how both~$\mathcal{A}_\mathrm{NH}$ and~$\mathcal{A}_\mathrm{HI}$ can be used to train~$\mathcal{F}_\theta$ for HLC-only.
The input signal~$x$ and the compensated output signal~$\hat{y}=\mathcal{F}_\theta(x, \Delta^\mathrm{dB})$ are processed by~$\mathcal{A}_\mathrm{NH}$ and~$\mathcal{A}_\mathrm{HI}$, respectively.
Similar to NR-only, the two auditory-model outputs are compared using a loss function~$\ell$.
The training objective can be expressed as
\begin{equation}
  \label{eq:loss_hlc}
  \mathcal{L}_\mathrm{HLC} = \ell \Bigl( \mathcal{A}_\mathrm{HI} \bigl( \mathcal{F}_\theta(x, \Delta^\mathrm{dB}), \Delta^\mathrm{dB} \bigr), \mathcal{A}_\mathrm{NH}(x) \Bigr).
\end{equation}
This effectively trains~$\mathcal{F}_\theta$ to process the input signal~$x$ such that the output of the HI auditory model~$\mathcal{A}_\mathrm{HI}$ is restored toward that of the NH auditory model~$\mathcal{A}_\mathrm{NH}$.

Previous studies in this category can be split into two groups: those that have used differentiable auditory models for~$\mathcal{A}_\mathrm{NH}$ and~$\mathcal{A}_\mathrm{HI}$~\cite{tu2021dhasp,drgas2024dynamic}, and those that have trained emulators beforehand to approximate nondifferentiable models of the auditory periphery~\cite{drakopoulos2022differentiable,drakopoulos2023neural,drakopoulos2023dnn,wen2025dconnear,leer2025hearing}.
They can be further categorized based on whether~$\mathcal{F}_\theta$ takes the audiogram~$\Delta^\mathrm{dB}$ as an additional input~\cite{drakopoulos2023dnn,drgas2024dynamic}, allowing personalization without retraining, or not~\cite{drakopoulos2022differentiable,drakopoulos2023neural,wen2025dconnear,leer2025hearing,tu2021dhasp}.
When~$\mathcal{F}_\theta$ does not take~$\Delta^\mathrm{dB}$ as an input, a separate DNN must be trained for each listener profile.
In~\cite{drakopoulos2023dnn}, even though~$\mathcal{F}_\theta$ took~$\Delta^\mathrm{dB}$ as an input, the auditory-model emulator~$\mathcal{A}_\mathrm{HI}$ was trained for individual listener profiles using transfer learning.

The input signal~$x$ can be clean~\cite{drakopoulos2022differentiable,drakopoulos2023neural,drakopoulos2023dnn,wen2025dconnear,tu2021dhasp,drgas2024dynamic} or noisy~\cite{leer2025hearing}.
Studies that trained with clean speech but evaluated under noisy conditions showed that the system can restore the auditory-model response to the unprocessed noisy speech~\cite{drakopoulos2023neural,drakopoulos2023dnn,wen2025dconnear}.
This suggests that the system learns to amplify all input sounds and does not perform zero-shot denoising, despite only seeing clean speech during training.
Therefore, it may be preferable to include noisy speech during training to diversify the observed acoustic conditions and further improve the performance in noise~\cite{leer2025hearing}.

\subsection{Joint NR and HLC}
\label{sec:previous_nr_hlc}

\noindent Figure~\ref{fig:previous_nr_hlc} describes how~$\mathcal{F}_\theta$ can be trained for joint NR and HLC.
This is similar to HLC-only, except that the input signal~$x$ to~$\mathcal{F}_\theta$ is noisy speech, and~$\mathcal{A}_\mathrm{NH}$ processes the corresponding clean speech~$y$ instead of~$x$.
The training objective is
\begin{equation}
  \label{eq:loss_nr_hlc}
  \mathcal{L}_\mathrm{NRHLC} = \ell \Bigl( \mathcal{A}_\mathrm{HI} \bigl( \mathcal{F}_\theta(x, \Delta^\mathrm{dB}), \Delta^\mathrm{dB} \bigr), \mathcal{A}_\mathrm{NH}(y) \Bigr).
\end{equation}
This approach was adopted by two studies~\cite{leer2025hearing,tu2021optimising}.
It results in a single DNN performing both NR and HLC simultaneously, which may be more efficient than designing two systems independently and applying them sequentially.
However, this comes at the cost of flexibility, since NR and HLC cannot be independently adjusted.
Moreover, since both tasks are optimized through the same training objective, it is unclear which task drives the optimization, and whether the DNN favors one over the other.
Additionally, these studies did not actually provide the audiogram~$\Delta^\mathrm{dB}$ as an input to~$\mathcal{F}_\theta$, and thus trained a separate DNN for each listener profile.
Therefore, it remains unclear whether a single DNN can restore the output of an auditory model across multiple hearing-loss profiles while simultaneously providing NR.

\section{Proposed multi-task framework}
\label{sec:proposed}

\begin{figure}
  \centering
  \subfloat[Training]{
    \includegraphics{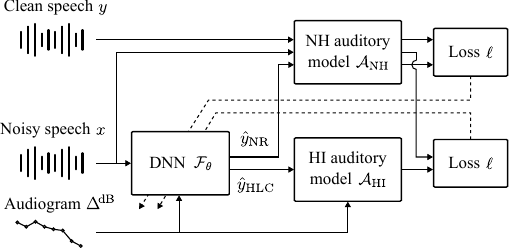}
    \label{fig:proposed_training}
  }\\
  \subfloat[Inference]{
    \includegraphics{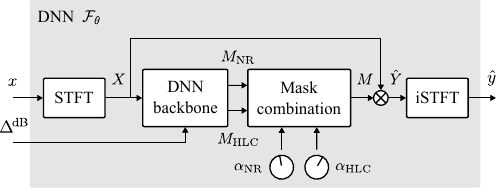}
    \label{fig:proposed_inference}
  }
  \caption{
    Proposed multi-task framework for adjustable joint NR and HLC using a single DNN and two auditory models.
    During training (a), the DNN predicts both a denoised and a compensated signal, and a distinct training objective is defined for each output.
    During inference (b), NR and HLC are adjusted independently by exponentiating each predicted TF mask.
  }
  \label{fig:proposed}
\end{figure}

\noindent To overcome the limitations of previous single-DNN approaches for joint NR and HLC, we propose the multi-task learning framework illustrated in Fig.~\ref{fig:proposed_training}.
In contrast to previous methods that produce a single output, the DNN~$\mathcal{F}_\theta$ is tasked with producing two separate outputs: a denoised signal~$\hat{y}_\mathrm{NR}$ and a compensated signal~$\hat{y}_\mathrm{HLC}$,
\begin{equation}
  \mathcal{F}_\theta(x, \Delta^\mathrm{dB}) = (\hat{y}_\mathrm{NR}, \hat{y}_\mathrm{HLC}).
\end{equation}
This allows defining distinct training objectives for NR and HLC, and in turn balancing the two tasks during training.
More specifically, we use the same objectives as in~\eqref{eq:loss_nr} and~\eqref{eq:loss_hlc},
\begin{align}
  \mathcal{L}_\mathrm{NR} &= \ell \bigl( \mathcal{A}_\mathrm{NH} ( \hat{y}_\mathrm{NR} ), \mathcal{A}_\mathrm{NH}(y) \bigr), \\
  \mathcal{L}_\mathrm{HLC} &= \ell \bigl( \mathcal{A}_\mathrm{HI} ( \hat{y}_\mathrm{HLC}, \Delta^\mathrm{dB} ), \mathcal{A}_\mathrm{NH}(x) \bigr).
\end{align}
In typical multi-task learning, the individual training objectives are combined using a weighted sum.
The weights remain fixed during training, and are commonly selected via a grid search.
However, grid searches can be time-consuming, and the results can be sensitive to the choice of weights.
Moreover, the optimal weights may change during training, and thus using fixed weights may not be ideal.
One solution for automatically adjusting the weights during training involves modeling the target observations as isotropic Gaussian distributions whose parameters are predicted by the DNN, and maximizing the corresponding log-likelihood~\cite{kendall2018multi}.
In practice, this reduces to optimizing two additional parameters~$u_\mathrm{NR} = \log \sigma_\mathrm{NR}^2$ and~$u_\mathrm{HLC} = \log \sigma_\mathrm{HLC}^2$ along with the DNN parameters~$\theta$, where~$\sigma_\mathrm{NR}>0$ and~$\sigma_\mathrm{HLC}>0$ represent the homoscedastic uncertainty related to each task.
The final training objective is
\begin{equation}
  \label{eq:loss_c_nr_hlc}
  \mathcal{L}_\mathrm{CNRHLC} = \frac{\mathcal{L}_\mathrm{NR}}{e^{u_\mathrm{NR}}} + u_\mathrm{NR} + \frac{\mathcal{L}_\mathrm{HLC}}{e^{u_\mathrm{HLC}}} + u_\mathrm{HLC}.
\end{equation}
The uncertainties~$\sigma_\mathrm{NR}$ and~$\sigma_\mathrm{HLC}$ are task-specific.
They are not data-dependent and do not measure DNN convergence.
Therefore, each loss term is weighted down if the related uncertainty is high.
Concurrently, these uncertainties are encouraged to be as small as possible, since their logarithms appear as additive terms penalizing the final loss.

To combine the two tasks and define a single output signal during inference, we consider a DNN backbone predicting two real- or complex-valued masks~$M_\mathrm{NR}$ and~$M_\mathrm{HLC}$ in the short-time Fourier transform (STFT) domain,
\begin{equation}
  \hat{Y}_\mathrm{NR} = M_\mathrm{NR} \odot X, \quad \hat{Y}_\mathrm{HLC} = M_\mathrm{HLC} \odot X,
\end{equation}
where~$X$, $\hat{Y}_\mathrm{NR}$, and~$\hat{Y}_\mathrm{HLC}$ are the STFTs of~$x$, $\hat{y}_\mathrm{NR}$, and~$\hat{y}_\mathrm{HLC}$, respectively, and~$\odot$ is the element-wise multiplication.
During inference, the predicted masks are independently exponentiated with two parameters~$\alpha_\mathrm{NR} \in [0, 1]$ and~$\alpha_\mathrm{HLC} \in [0, 1]$.
More specifically, for each TF unit~$(t, f)$, the exponentiated masks~$\tilde{M}_\mathrm{NR}$ and~$\tilde{M}_\mathrm{HLC}$ are computed as
\begin{equation}
  \tilde{M}_\mathrm{NR}^{t, f} = \bigl( M_\mathrm{NR}^{t, f} \bigr)^{\alpha_\mathrm{NR}}, \quad \tilde{M}_\mathrm{HLC}^{t, f} = \bigl( M_\mathrm{HLC}^{t, f} \bigr)^{\alpha_\mathrm{HLC}}.
\end{equation}
This exponentiation linearly scales the magnitude in dB and, for complex-valued masks, also linearly scales the phase.
The two exponentiated masks are then combined by multiplying them together element-wise.
This combination is conceptually natural because each mask is multiplicative in nature, and thus multiplying the two masks is equivalent to chaining the two operations.
Compared with a linear combination, our approach does not produce abrupt phase changes when the combined mask approaches zero, and ensures monotonic variations in magnitude and phase.
To allow gradual adjustment of NR for noise-only TF units where the predicted mask~$M_\mathrm{NR}$ may be close to zero, the magnitude of the combined mask is thresholded using an~$\alpha_\mathrm{NR}$-dependent minimum gain.
Limiting the attenuation also helps balance speech intelligibility and perceived quality~\cite{brons2012perceptual,borgstrom2021speech,johnson2024ideal}.
The final mask~$M$ is computed for each TF unit~$(t, f)$ as
\begin{gather}
  | M^{t, f} | = \max \bigl( | \tilde{M}_\mathrm{NR}^{t, f} | | \tilde{M}_\mathrm{HLC}^{t, f} |, {G_{\min}}^{\alpha_\mathrm{NR}} \bigr), \label{eq:combined_mask} \\
  \angle M^{t, f} = \angle \tilde{M}_\mathrm{NR}^{t, f} + \angle \tilde{M}_\mathrm{HLC}^{t, f}, \\
  M^{t, f} = | M^{t, f} | e^{j \angle M^{t, f}},
\end{gather}
where~$G_{\min}$ is the desired minimum gain when~$\alpha_\mathrm{NR} = 1$.
The amplification can also be limited using a maximum gain~$G_{\max}$.
However, such a constraint is not evaluated in the present study.
The mask~$M$ is then applied to the noisy STFT~$X$,
\begin{equation}
  \hat{Y} = M \odot X.
\end{equation}
The final time-domain output~$\hat{y}$ is obtained by computing the inverse STFT (iSTFT) of~$\hat{Y}$.
When~$\alpha_\mathrm{NR}=\alpha_\mathrm{HLC}=1$, maximum NR and HLC are applied according to the predicted masks.
Conversely, when~$\alpha_\mathrm{NR}=\alpha_\mathrm{HLC}=0$, no NR or HLC is applied, and the output~$\hat{y}$ is identical to the input~$x$.

Several speech enhancement studies share similarities with our framework.
In~\cite{cao2022cmgan,abdulatif2024cmgan,li2022glance,zhao2023d2former}, separate neural network modules predict masking and mapping components.
These components are combined to form a clean speech estimate, from which the training objective is computed.
While multiple terms are sometimes used in the training objective, they all regress toward the same clean speech.
This contrasts with our approach, where two distinct output signals are defined, each with its own training objective, and their combination occurs only at inference time.
In~\cite{li2021two,lu2023mp,lu2025explicit}, intermediate training objectives are defined, but they still regress toward the same clean reference.
In~\cite{xu2015multi,sun2017multiple,zhou2021masks,shi2020spectrograms}, multiple supervised outputs are combined after training, similar to our approach.
However, each target is an alternative representation of the same clean reference, with additional targets acting as regularizers to improve the final estimate.
In~\cite{shi2020spectrograms}, the fusion weights are predicted by a second network trained to minimize the distance to the clean reference, and in~\cite{shi2021spectrograms} the fused front end is fine-tuned jointly with a speech recognizer.
More generally, all these approaches consider a single enhancement task with a single clean target, whereas our work addresses two distinct tasks, NR and HLC, with separate task-specific targets.

\section{Experimental setup}
\label{sec:setup}

\subsection{DNN backbone and STFT parameters}
\label{sec:setup_dnn}

\noindent The DNN backbone is based on the \textit{band-split recurrent neural network} (BSRNN)~\cite{luo2023music}.
BSRNN achieves state-of-the-art results in speech enhancement~\cite{yu2023efficient,yu2023high} and offers a strong balance between computational cost and performance~\cite{zhang2024beyond}.
A BSRNN-based system also achieved first place in the recent URGENT 2025 Speech Enhancement Challenge~\cite{sun2025scaling}.

\begin{figure}
  \centering
  \includegraphics[width=0.98\linewidth]{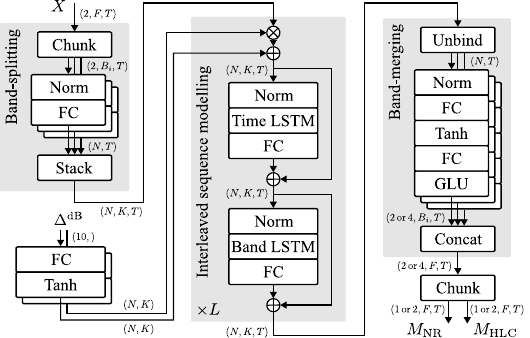}
  \caption{Overview of the BSRNN-based DNN backbone.}
  \label{fig:bsrnn}
\end{figure}

Figure~\ref{fig:bsrnn} provides an overview of the DNN backbone architecture.
The real and imaginary parts of the input STFT~$X$ are stacked along the channel dimension to form an input tensor of shape~$(2, F, T)$, where~$F$ denotes the number of frequency bins and~$T$ the number of time frames.
The band-split module projects~$K$ frequency bands with increasing width~$\{B_i\}_{i=1}^K$, where~$\sum_{i=1}^K B_i = F$, onto a fixed number of channels~$N$ using band-specific fully connected (FC) layers.
Each FC layer is preceded by global layer normalization (gLN)~\cite{luo2019conv}, which aggregates statistics over all channels, time frames, and frequency bins within the current band~\cite{wu2018group}.
Interleaved sequence modeling across time frames and frequency bands is performed using residual bidirectional long short-term memory (LSTM) blocks in~$L$ layers.
Each LSTM is preceded by gLN and followed by an FC layer.
If the DNN performs HLC, features are extracted from the audiogram using band-specific FC layers and a tanh activation function, and are included before each sequence modeling layer using feature-wise linear modulation (FiLM)~\cite{perez2018film}.
The band-merging module projects each frequency band back to the original STFT resolution using band-specific multilayer perceptrons (MLPs).
Each MLP is preceded by gLN, which aggregates statistics over all channels and time frames.
Each MLP uses a tanh activation function after each hidden layer, and a gated linear unit (GLU)~\cite{dauphin2017language} after the output layer.
The number of output channels in each MLP is doubled when predicting complex-valued masks, and further doubled when predicting separate masks~$M_\mathrm{NR}$ and~$M_\mathrm{HLC}$ for adjustable NR and HLC as described in Fig.~\ref{fig:proposed}.

The STFT uses a Hann window, a frame length of \qty{32}{\milli\second}, and a hop size of \qty{16}{\milli\second}.
We use~$K = 32$ frequency bands with bounds uniformly spaced on a mel scale.
The number of channels is~$N = 64$, the number of layers is~$L = 6$, and the number of hidden channels in each LSTM direction is~$2 N = 128$.
Each MLP in the band-merging module has one hidden layer with~$4 N = 256$ channels.
The total number of trainable parameters ranges from \qty{3.3}{\mega\nothing} to \qty{3.7}{\mega\nothing}, depending on whether the DNN predicts real- or complex-valued masks, takes the audiogram as additional input for HLC, and predicts one or two masks.
Because the architecture uses normalization across time and bidirectional LSTMs, the DNN is noncausal.
Moreover, the STFT and its inverse have an algorithmic latency of \qty{32}{\milli\second}, which is too high for direct real-time hearing-aid implementation.
Adapting the proposed framework to causal, low-latency processing is left for future work.

\subsection{Differentiable auditory model}
\label{sec:setup_am}

\begin{figure}
  \centering
  \vspace{-\baselineskip}
  \subfloat[Auditory-model stages]{
    \includegraphics[valign=c]{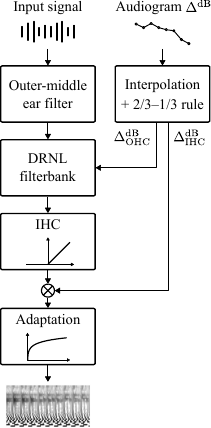}
    \label{fig:am}
  }\hfil
  \subfloat[Single DRNL filter]{
    \includegraphics[valign=c]{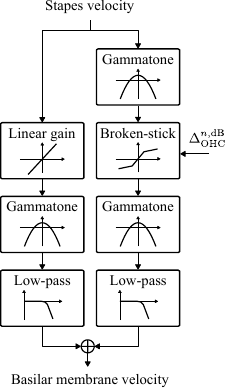}
    \vphantom{\includegraphics[valign=c]{pics/am.pdf}}
    \label{fig:drnl}
  }
  \caption{Overview of the proposed differentiable auditory model.}
  \label{fig:am_drnl}
\end{figure}

\noindent The differentiable auditory model used in this study is a simplified auditory front-end inspired by established models of auditory processing and masking~\cite{dau1996quantitative,dau1997modeling,jepsen2008computational,paulick2025computational}.
In particular, it follows the general structure of peripheral filtering, envelope extraction, and compressive processing that has been used successfully to account for a range of psychoacoustic data.
At the same time, the model is deliberately simplified to make it suitable for end-to-end optimization: all stages are differentiable, the computational complexity is reduced, and only the stages considered most relevant for the present joint NR and HLC objective are retained.
Figure~\ref{fig:am} gives an overview of the resulting model.
Speech signals are resampled from \qty{16}{\kilo\hertz} to the operating sampling rate of the auditory model~$f_s=\qty{24}{\kilo\hertz}$.

\subsubsection{Outer-middle ear filter~\texorpdfstring{\cite{lopez2001human}}{}}

The first stage is a 512-tap FIR filter modeling the combined transfer function of the outer and middle ear, transforming headphone-delivered sound pressure into stapes velocity at the cochlear base.
This stage can be implemented directly in a differentiable way without the need for approximation.

\subsubsection{Dual-resonance nonlinear (DRNL) filterbank~\texorpdfstring{\cite{lopez2001human}}{}}

The cochlear stage is implemented using a bank of~$N_\mathrm{CF}=49$ DRNL filters with center frequencies evenly spaced by~$\delta_\mathrm{ERB}=0.67$ on the equivalent rectangular bandwidth (ERB)-rate scale between~$\mathrm{CF}_\mathrm{min}=\qty{80}{\hertz}$ and~$\mathrm{CF}_\mathrm{max}=\qty{9854}{\hertz}$.
The choice of the number of center frequencies and their range is motivated in Section~\ref{sec:results_ltag}.
This stage models the frequency selectivity of the basilar membrane along the cochlea as well as the level-dependent amplification provided by outer hair cells (OHCs).
Figure~\ref{fig:drnl} shows the structure of a single DRNL filter.
Each filter consists of a linear and a nonlinear path in parallel.
The linear path includes a linear gain, a gammatone filter, and a low-pass Butterworth filter.
The nonlinear path consists of a first gammatone filter, a broken-stick nonlinearity, a second gammatone filter, and a low-pass Butterworth filter.
The broken-stick nonlinearity is defined as
\begin{equation}
  h = \sgn(g) \cdot \min ( a_n |g| \Delta_\mathrm{OHC}^{n}, b_n |g|^{c_n} ),
\end{equation}
where~$g$ and~$h$ are the input and output signals, respectively, $n \in \{1, \ldots, N_\mathrm{CF}\}$ is the DRNL filter index, $a_n$, $b_n$, and~$c_n$ are parameters controlling the shape of the nonlinearity, and~$\Delta_\mathrm{OHC}^{n}$ is the OHC-induced hearing loss on a linear scale.
The outputs of the linear and nonlinear paths are summed to form the DRNL output.
The complete set of DRNL parameters is taken from~\cite{lopez2001human}.
To accelerate training, the gammatone and low-pass filters are implemented as 512-tap FIR filters rather than infinite impulse response (IIR) filters.
The broken-stick nonlinearity is differentiable everywhere except at~$g = 0$, and therefore~$|g|$ is thresholded using a very small value.

\subsubsection{Inner hair cell (IHC) transduction}

The next stage models the transformation from basilar-membrane vibrations to a neural representation by the IHCs.
In the present model, this stage is implemented as half-wave rectification followed by listener-specific gain adjustment.
In contrast to more detailed peripheral models, no additional low-pass filtering is applied here.
Preliminary experiments showed that including such a low-pass stage tended to produce artifacts in the TF masks estimated by the DNN.
The IHC-induced hearing loss for the~$n$-th center frequency, denoted by~$\Delta_\mathrm{IHC}^{n}$, is applied as a gain function at the output of this stage.

\subsubsection{Compression stage}

Classical auditory models for masking and modulation processing often include a cascade of adaptation loops to account for neural adaptation and dynamic compression~\cite{dau1996quantitative}.
However, the recursive nature of the adaptation loops makes it difficult to implement them efficiently in a differentiable framework.
Therefore, in the present work, this stage is replaced by an instantaneous static log-compression,
\begin{equation}
  v = \log(1 + u / u_0),
\end{equation}
where~$u$ and~$v$ are the input and output signals, respectively, and~$u_0 = 1 \cdot 10^{-5}$ for all center frequencies.

In the original auditory model from~\cite{dau1997modeling}, the compressed output is followed by a modulation filterbank that represents modulation-frequency selective processing.
Preliminary experiments indicated that including a modulation filterbank degraded NR performance, presumably because it removes signal components outside the modulation-frequency range that are nevertheless informative for separating speech from noise.
Therefore, in the present study, such a stage is omitted.

The OHC- and IHC-induced hearing losses~$\Delta_\mathrm{OHC}^{n}$ and~$\Delta_\mathrm{IHC}^{n}$ are estimated by first linearly interpolating the audiogram thresholds in dB~$\Delta^\mathrm{dB}$ from the audiometric frequencies to the DRNL center frequencies on a logarithmic frequency axis.
Following previous studies~\cite{lopez2012behavioral,zaar2022predicting,relano2023evaluating}, each interpolated threshold~$\Delta_\mathrm{interp.}^{n,\mathrm{dB}}$ is then split into OHC- and IHC-induced hearing losses using the following 2/3--1/3 rule:
\begin{align}
  \Delta_\mathrm{OHC}^{n,\mathrm{dB}} &= \min \biggl( \frac{2}{3} \Delta_\mathrm{interp.}^{n,\mathrm{dB}}, \Delta_\mathrm{OHC,max}^{n,\mathrm{dB}} \biggr), \\
  \Delta_\mathrm{IHC}^{n,\mathrm{dB}} &= \Delta_\mathrm{interp.}^{n,\mathrm{dB}} - \Delta_\mathrm{OHC}^{n,\mathrm{dB}},
\end{align}
where~$\Delta_\mathrm{OHC,max}^{n,\mathrm{dB}}$ denotes the maximum OHC-induced hearing loss that can be represented by the~$n$-th DRNL filter.

\subsection{Datasets}
\label{sec:setup_datasets}

\noindent DNNs are trained with noisy and reverberant speech sampled at \qty{16}{\kilo\hertz}.
Each acoustic scene is simulated by placing one speech source and one to three noise sources in the same room.
Clean speech utterances are selected from DNS5~\cite{dubey2024icassp}, LibriSpeech~\cite{panayotov2015librispeech}, MLS~\cite{pratap2020mls}, VCTK~\cite{veaux2013voice}, and EARS~\cite{richter2024ears}.
Noise segments are selected from DNS5~\cite{dubey2024icassp}, WHAM!~\cite{wichern2019wham}, FSD50K~\cite{fonseca2022fsd50k}, and FMA~\cite{defferrard2017fma}.
The total amount of available speech and noise is \qtylist{1713;541}{\hour}, respectively.
The room impulse responses (RIRs) are simulated using a fast random approximation of the image-source method proposed in~\cite{luo2024fast}.
The room size is uniformly selected between \qtyproduct{3x3x2.5}{\metre} and \qtyproduct{10x10x4}{\metre}.
The reverberation time~$T_{60}$ is uniformly selected between \qtylist{0.1;0.7}{\second}.
The level of each reverberant noise source is uniformly selected within a \qty{10}{\decibel} range to randomize the relative levels of the noise sources.
The SNR between the reverberant speech and the sum of the reverberant noise sources is uniformly selected between \qtylist{-5;15}{\decibel}.
The level of the total reverberant mixture is uniformly selected between \qtylist{65;85}{\decibel} SPL using the same convention as in the Auditory Modeling Toolbox~\cite{sondergaard2013auditory,majdak2022amt}, i.e.\ a root mean square (RMS) value of 1 corresponds to \qty{93.98}{\decibel} SPL.
When training for NR, early speech reflections are included in the target signal~$y$ using a reflection boundary of \qty{50}{\milli\second} as suggested in previous studies~\cite{bradley2003importance,roman2013speech}.
Therefore, systems trained for NR also perform dereverberation.
Acoustic scenes are generated on-the-fly during training to maximize data diversity and improve generalization~\cite{zeghidour2021wavesplit,choi2022empirical}.
For testing, \num{1000} scenes are generated using unseen speech utterances, noise segments, and RIRs from Clarity~\cite{graetzer2022dataset}, TUT~\cite{mesaros2016tut}, and DNS5~\cite{dubey2024icassp}, respectively.
All corpora used in this study are publicly available.

\subsection{Training}
\label{sec:setup_training}

\noindent All DNNs are trained with 4-\unit{\second}-long scenes for 300 epochs, with \num{7500} scenes per epoch.
We use a batch size of 24 and the Adam optimizer~\cite{kingma2015adam} with an initial learning rate of~$1 \cdot 10^{-3}$.
The learning rate is multiplied by 0.99 after each epoch.
Gradients are clipped with a maximum~$L_2$ norm of 5.
Training takes approximately \qty{36}{\hour} on a single \qty{48}{\giga\byte} graphics processing unit.

\subsection{Audiograms}
\label{sec:setup_audiograms}

\begin{figure}
  \centering
  \includegraphics[scale=0.66]{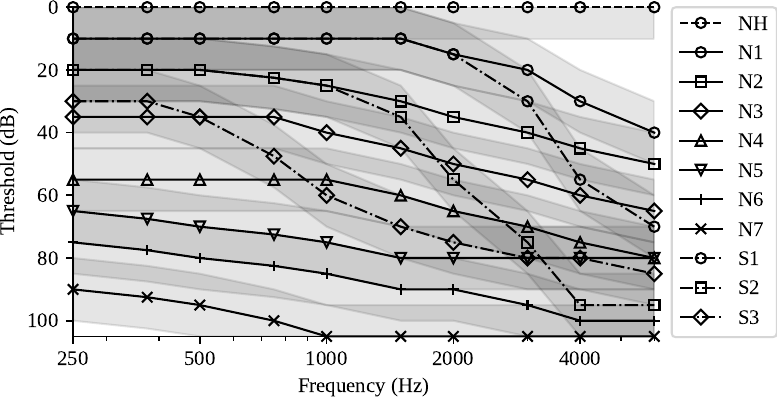}
  \caption{Standard audiograms. Shaded areas indicate the range of thresholds covered by the \qty{10}{\decibel} random jitter during training. The randomized thresholds are then constrained to the range [0,\,105] \unit{\decibel}.}
  \label{fig:audiograms}
\end{figure}

\noindent We consider the set of 10 standard audiograms from~\cite{bisgaard2010standard}, which we extend with an NH audiogram for which all thresholds are set to \qty{0}{\decibel}.
When training for HLC, an audiogram~$\Delta^\mathrm{dB}$ is uniformly selected from the 11 profiles for each training scene.
To increase the diversity of audiograms observed by the DNN during training, a uniformly distributed random jitter between \qtylist{-10;10}{\decibel} is applied independently to each threshold, after which thresholds are constrained to the range [0,\,105] \unit{\decibel}.
The resulting distribution of audiograms used during training is illustrated in Fig.~\ref{fig:audiograms}.
Since the perturbation is applied independently at each audiometric frequency, the training audiograms are not restricted to shifted versions of the standard profiles, and contain local frequency-dependent variations around the underlying flat and high-frequency-sloping configurations.
During testing, each scene is processed for each of the 11 standard audiograms without applying any jitter.
Audiometric frequencies are fixed to \qtylist{250;375;500;750;1000;1500;2000;3000;4000;6000}{\hertz}.
Each threshold in dB is divided by 100 at the input of the DNN to normalize the input scale.
Only the thresholds are provided as input to the DNN, i.e.\ the frequencies are discarded, although including the frequencies may allow the DNN to support audiograms measured at arbitrary frequencies.

\subsection{Objective metrics}
\label{sec:setup_metrics}

\noindent Systems are evaluated for NR and joint NR and HLC separately.
NR performance is evaluated using perceptual evaluation of speech quality (PESQ)~\cite{rix2001perceptual}, extended short-term objective intelligibility (ESTOI)~\cite{jensen2016algorithm}, and signal-to-distortion ratio (SDR).
For these metrics, an NH audiogram is provided to systems capable of performing HLC.
Joint NR and HLC performance is evaluated using hearing-aid speech perception index (HASPI)~\cite{kates2014hearing} and hearing-aid speech quality index (HASQI)~\cite{kates2010hearing}.
For these metrics, each test scene is processed for each of the 11 standard audiograms, and results are averaged over all audiograms unless stated otherwise.
For all metrics, the reference signal~$y$ only includes the early speech reflections as described in Section~\ref{sec:setup_datasets}.
We use the HASPI and HASQI implementations from the Clarity Challenge toolkit~\cite{barker2023pyclarity}.

To evaluate the ability of the system to restore the output of the HI auditory model~$\mathcal{A}_\mathrm{HI}$ to that of the NH auditory model~$\mathcal{A}_\mathrm{NH}$, we also report the normalized RMS error (NRMSE) between the NH population response~$r_\mathrm{p}$ to the reference speech~$y$ and the HI population response~$\hat{r}_\mathrm{p}$ to the processed speech~$\hat{y}$, similar to~\cite{drakopoulos2023neural,drakopoulos2023dnn,wen2025dconnear,wen2025individualized},
\begin{equation}
  r_\mathrm{p}^{m} = \sum_{n=1}^{N_\mathrm{CF}} \mathcal{A}^{n, m}_\mathrm{NH} ( y ), \quad \hat{r}_\mathrm{p}^{m}= \sum_{n=1}^{N_\mathrm{CF}} \mathcal{A}^{n, m}_\mathrm{HI} ( \hat{y}, \Delta^\mathrm{dB} ),
\end{equation}
where~$m$ is the time sample index.
NRMSE is computed as
\begin{equation}
  \mathrm{NRMSE} = \frac{1}{\max\limits_{m \in \{ 1, \ldots, M\}} {r_\mathrm{p}^{m}}} \sqrt{ \frac{1}{M} \sum_{m=1}^{M} \bigl( r_\mathrm{p}^{m} - \hat{r}_\mathrm{p}^{m} \bigr)^2 },
\end{equation}
where~$M$ is the total number of time samples.
We stress that NRMSE specifically evaluates auditory-model output restoration, and is intended to verify the behavior of the learned HLC rather than to predict perceptual benefit.
Similar to HASPI and HASQI, NRMSE is averaged over all standard audiograms.

\subsection{Compared systems}
\label{sec:setup_systems}

\noindent The following BSRNN-based DNNs are trained using the differentiable auditory model proposed in Section~\ref{sec:setup_am}:
\begin{itemize}
  \item \textit{DNN-NR}: baseline predicting one mask and trained for NR-only using~\eqref{eq:loss_nr} as the training objective.
  \item \textit{DNN-HLC}: baseline predicting one mask and trained for HLC-only using~\eqref{eq:loss_hlc} as the training objective.
  \item \textit{DNN-NRHLC}: baseline predicting one mask and trained for joint NR and HLC using~\eqref{eq:loss_nr_hlc} as the training objective.
  \item \textit{DNN-CNRHLC}: proposed solution predicting two masks and trained for adjustable joint NR and HLC using~\eqref{eq:loss_c_nr_hlc}.
\end{itemize}
An initial experiment compares these systems when predicting either real- or complex-valued masks and using either the mean squared error (MSE) or the mean absolute error (MAE) as the loss function~$\ell$.
We then investigate the effect of adjusting NR and HLC during inference by varying~$\alpha_\mathrm{NR}$ and~$\alpha_\mathrm{HLC}$ for DNN-CNRHLC as described in Section~\ref{sec:proposed}.
Finally, the selected systems are compared with the traditional hearing-aid prescription NAL-R~\cite{byrne1986national}, following previous work~\cite{leer2025hearing,drakopoulos2023dnn,drakopoulos2022differentiable,tu2021dhasp,tu2021optimising}.
We also investigate applying systems sequentially.
For example, DNN-NR\,+\,DNN-HLC denotes a system that processes the noisy input signal with DNN-NR for NR first, and then processes the denoised signal with DNN-HLC for HLC.
Similarly, DNN-NR\,+\,NAL-R denotes a system that chains DNN-NR for NR and NAL-R for HLC.
An additional NR-only baseline DNN-SDR is trained with SDR as the training objective, as commonly done in speech enhancement.
We use SDR instead of scale-invariant signal-to-distortion ratio (SI-SDR)~\cite{leroux2019sdr} so that the DNN preserves the level of the target signal as in previous studies~\cite{kinoshita2020improving,tu2021two,zmolikova2021but}.
The magnitude of the predicted mask is thresholded using~$G_{\min}$ for all DNNs, except DNN-CNRHLC for which the combined mask is thresholded using an~$\alpha_\mathrm{NR}$-dependent minimum gain as in~\eqref{eq:combined_mask}.
Following previous studies~\cite{borgstrom2021speech,johnson2024ideal}, we set~$G_{\min}$ to \qty{-25}{\decibel}.
To investigate the potential of adjusting NR and HLC during inference, we also report results for DNN-CNRHLC$^*$, for which the best~$\alpha_\mathrm{NR}$ and~$\alpha_\mathrm{HLC}$ are selected for each listener, scene, and metric post hoc, instead of fixing them across all configurations as for DNN-CNRHLC.
We stress that DNN-CNRHLC$^*$ is an oracle system providing an upper bound on the objective metrics potentially achievable through adaptive adjustment of NR and HLC, and does not represent a practically applicable strategy.
We use the NAL-R implementation from the Clarity Challenge toolkit~\cite{barker2023pyclarity}.

\section{Results}
\label{sec:results}

\subsection{Real vs.\ complex masks and MSE vs.\ MAE loss}
\label{sec:results_barplot}

\begin{figure}
  \centering
  \includegraphics[scale=0.68]{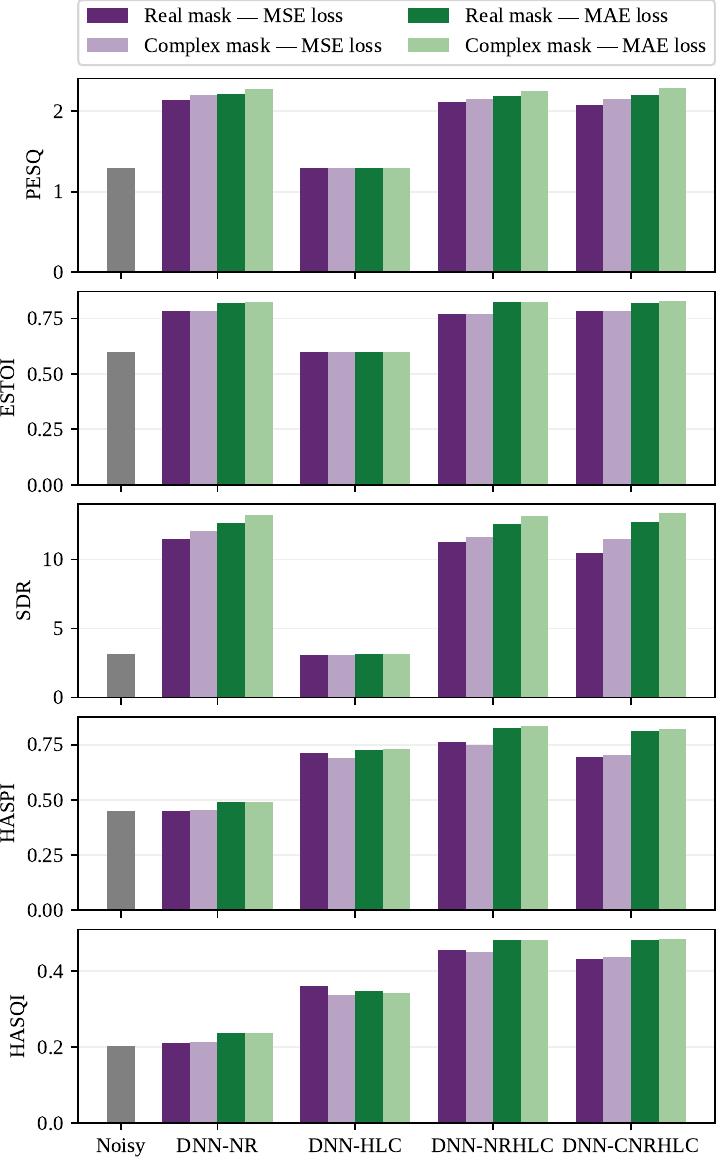}
  \caption{
    Objective metrics for real- or complex-valued masks and MSE or MAE loss.
    For PESQ, ESTOI, and SDR, an NH audiogram is provided to systems providing HLC.
    For HASPI and HASQI, results are averaged over all audiograms.
    Both~$\alpha_\mathrm{NR}$ and~$\alpha_\mathrm{HLC}$ are set to 1 for DNN-CNRHLC.
  }
  \label{fig:results_barplot}
\end{figure}

\noindent Figure~\ref{fig:results_barplot} shows the average objective metrics for the systems trained with the proposed differentiable auditory model when using either real- or complex-valued masks and either MSE or MAE as the loss function~$\ell$.
For DNN-CNRHLC, both~$\alpha_\mathrm{NR}$ and~$\alpha_\mathrm{HLC}$ are set to 1.
DNN-NRHLC and DNN-CNRHLC, which are trained for both NR and HLC, show similar NR performance to DNN-NR, which is trained for NR-only, as reflected by PESQ, ESTOI, and SDR.
This suggests that the same system can be optimized for both NR and HLC without compromising NR metrics.
One possible explanation is that the internal representations learned by the system are useful for both tasks.
DNN-HLC shows similar NR performance to the unprocessed noisy signal as expected, since an NH audiogram is provided to the system, and thus no HLC is applied.
This behavior is not explicitly enforced by the system architecture; it emerges from the proposed optimization framework.

DNN-NR achieves similar HASPI and HASQI to the unprocessed noisy signal, which suggests that NR-only is not sufficient to improve speech intelligibility and quality for HI profiles.
Providing HLC-only substantially improves both HASPI and HASQI, as shown by DNN-HLC.
Joint NR and HLC further improves HASQI over HLC-only, as shown by DNN-NRHLC and DNN-CNRHLC, while HASPI improves consistently for complex-valued masks only.
More generally, the best objective metrics are most commonly achieved when using a complex-valued mask and MAE loss.
Therefore, in the following sections, we use complex-valued masks and MAE loss for all DNN-based systems.

\begin{figure*}
  \centering
  \includegraphics[scale=0.6]{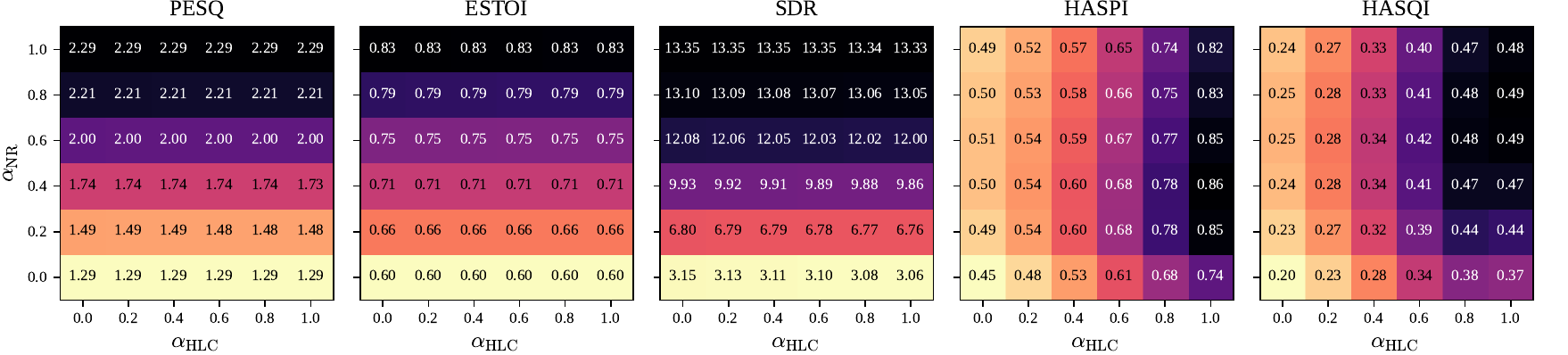}
  \caption{
    Objective metrics as a function of~$\alpha_\mathrm{NR}$ and~$\alpha_\mathrm{HLC}$.
    For PESQ, ESTOI, and SDR, an NH audiogram is provided to the DNN.
    For HASPI and HASQI, results are averaged over all audiograms.
  }
  \label{fig:results_heatmap}
\end{figure*}

\begin{figure}
  \centering
  \includegraphics[scale=0.6]{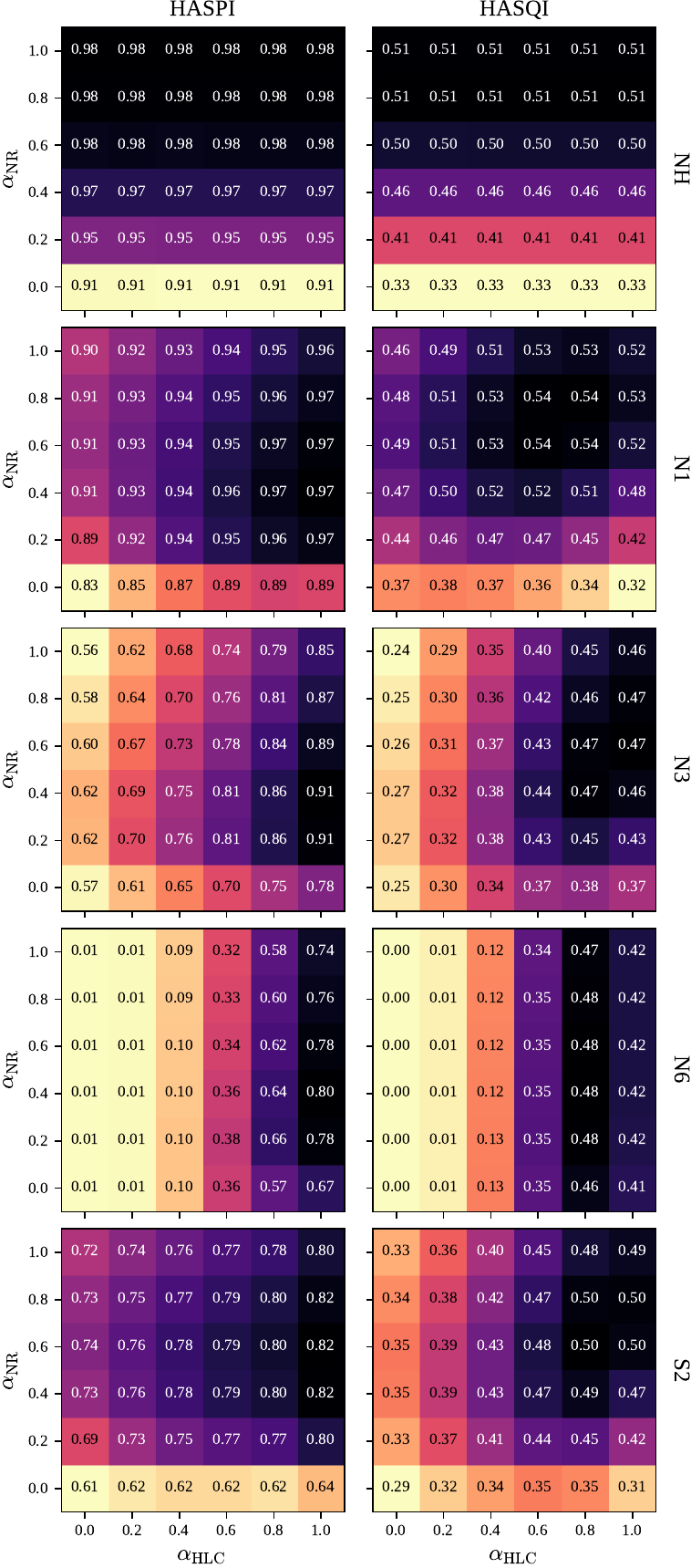}
  \caption{
    HASPI and HASQI as a function of~$\alpha_\mathrm{NR}$ and~$\alpha_\mathrm{HLC}$ for the NH audiogram and HI audiograms N1, N3, N6, and S2.
  }
  \label{fig:results_heatmap_2}
\end{figure}

\subsection{Influence of \texorpdfstring{$\alpha_\mathrm{NR}$ and~$\alpha_\mathrm{HLC}$}{alpha NR and alpha HLC}}
\label{sec:results_heatmap}

\noindent Figure~\ref{fig:results_heatmap} shows the average objective metrics as a function of~$\alpha_\mathrm{NR}$ and~$\alpha_\mathrm{HLC}$ for DNN-CNRHLC.
Each exponent is varied between 0 and 1 with a step size of 0.1, while the results are reported in steps of 0.2 for clarity.
For PESQ, ESTOI, and SDR, the results improve as~$\alpha_\mathrm{NR}$ increases as expected.
Meanwhile, they remain constant as~$\alpha_\mathrm{HLC}$ varies, which is also expected since an NH audiogram is provided to the system, and thus there is no hearing loss to compensate for.
Note that, similar to DNN-HLC in Section~\ref{sec:results_barplot}, this behavior is not explicitly enforced, but naturally emerges from the training procedure.
For HASPI, the best results are achieved for~$\alpha_\mathrm{NR}=0.4$ and~$\alpha_\mathrm{HLC}=1$, while for HASQI, they are achieved for~$\alpha_\mathrm{NR}=0.8$ and~$\alpha_\mathrm{HLC}=1$.
To select a single pair of values for the following experiments, we generate a validation dataset consisting of 100 noisy speech mixtures unseen during training.
The clean speech utterances and noise segments are drawn from the same corpora used for training, but the dataset generation is initialized with a different random seed, leading to different combinations of speech, noise, RIR, SNR, and level.
The pair of values achieving the best average ranking across HASPI and HASQI on this validation dataset is~$\alpha_\mathrm{NR}=0.7$ and~$\alpha_\mathrm{HLC}=1$.
We find that this pair of values also achieves the best average ranking across HASPI and HASQI for the test dataset, supporting the robustness of this parameter choice across acoustic conditions.

Figure~\ref{fig:results_heatmap_2} shows the HASPI and HASQI results as a function of~$\alpha_\mathrm{NR}$ and~$\alpha_\mathrm{HLC}$ for specific listener profiles, namely the NH audiogram and HI audiograms N1, N3, N6, and S2.
For the NH audiogram, the results remain constant as~$\alpha_\mathrm{HLC}$ is varied, since there is no hearing loss to compensate for.
For the HI audiograms, the sensitivity of HASPI and HASQI to~$\alpha_\mathrm{NR}$ and~$\alpha_\mathrm{HLC}$ varies considerably across profiles, and the optimal pair of values varies moderately.
This highlights the benefit of the ability to adjust NR and HLC at inference time, as~$\alpha_\mathrm{NR}$ and~$\alpha_\mathrm{HLC}$ can be tuned for each listener without retraining.

\subsection{Average results for all systems}
\label{sec:results_mean}

\begin{table}
\caption{Objective metrics for all DNN-based systems, NAL-R, and various combinations. For PESQ, ESTOI, and SDR, an NH audiogram is provided to systems providing HLC. For HASPI, HASQI, and NRMSE, results are averaged over all audiograms. DNN-CNRHLC uses~$(\alpha_\mathrm{NR},\alpha_\mathrm{HLC}) = (1, 1)$ for PESQ, ESTOI, and SDR, and~$(0.7, 1)$ for HASPI, HASQI, and NRMSE. DNN-CNRHLC$^*$ uses optimal~$(\alpha_\mathrm{NR},\alpha_\mathrm{HLC})$ for each listener, scene, and metric, and is not practically applicable. Values in \textbf{bold} indicate best results among DNN-based systems. Second best is \underline{underlined}.}
\label{tab:results}
\centering
\setlength{\tabcolsep}{4.7pt}
\begin{tabular}{lS[table-format=1.2]S[table-format=1.2]S[table-format=3.1]S[table-format=2.1]S[table-format=2.1]S[table-format=2.1]S[table-format=2.1]}
\toprule
& {\adjustbox{angle=90}{Parameters (M)}} & {\adjustbox{angle=90}{PESQ $\uparrow$}} & {\adjustbox{angle=90}{ESTOI (\%) $\uparrow$}} & {\adjustbox{angle=90}{SDR $\uparrow$}} & {\adjustbox{angle=90}{HASPI (\%) $\uparrow$}} & {\adjustbox{angle=90}{HASQI (\%) $\uparrow$}} & {\adjustbox{angle=90}{NRMSE (\%) $\downarrow$}} \\
\midrule
Noisy & {--} & {1.29} & {59.9} & {3.2} & {45.1} & {20.2} & {26.0} \\
Noisy\,+\,NAL-R & {--} & {1.29} & {59.9} & {3.2} & {70.7} & {57.2} & {26.0} \\
Clean\,+\,NAL-R & {--} & {4.64} & {100.0} & $\infty$ & {84.2} & {99.5} & {24.3} \\
\midrule
DNN-SDR & 3.42 & {\textbf{2.30}} & {\textbf{83.2}} & {\textbf{13.6}} & {50.1} & {24.5} & {25.7} \\
DNN-SDR\,+\,NAL-R & 3.42 & {\textbf{2.30}} & {\textbf{83.2}} & {\textbf{13.6}} & {74.5} & {\textbf{72.0}} & {25.7} \\
DNN-NR & 3.42 & {2.28} & {82.4} & {13.2} & {49.2} & {23.8} & {25.9} \\
DNN-NR\,+\,NAL-R & 3.42 & {2.28} & {82.4} & {13.2} & {72.7} & {\underline{70.7}} & {25.9} \\
DNN-HLC & 3.46 & {1.29} & {59.9} & {3.2} & {73.0} & {34.1} & {18.4} \\
DNN-NRHLC & 3.46 & {2.25} & {82.3} & {13.1} & {83.4} & {48.1} & {13.4} \\
DNN-CNRHLC & 3.73 & {\underline{2.29}} & {\underline{83.0}} & {\underline{13.3}} & {\underline{84.2}} & {49.0} & {\underline{13.1}} \\
DNN-CNRHLC$^*$ & 3.73 & {\underline{2.29}} & {\underline{83.0}} & {\underline{13.3}} & {\textbf{88.0}} & {51.0} & {\textbf{11.5}} \\
DNN-NR\,+\,DNN-HLC & 6.88 & {2.28} & {82.4} & {13.2} & {81.4} & {45.8} & {13.9} \\
\bottomrule
\end{tabular}
\end{table}

\noindent Table~\ref{tab:results} shows the average objective metrics for all DNN-based systems, NAL-R, and different combinations of systems.
The proposed DNN-CNRHLC achieves similar NR performance to DNN-SDR, as reflected by PESQ and ESTOI.
The output of the proposed differentiable auditory model is thus a suitable training target for NR.
Moreover, since DNN-CNRHLC is optimized for both NR and HLC, this confirms the observation from Section~\ref{sec:results_barplot} that the same system can be optimized for both tasks without compromising NR metrics.

The systems trained with the proposed differentiable auditory model for joint NR and HLC, namely DNN-NRHLC, DNN-CNRHLC, and DNN-NR\,+\,DNN-HLC, show considerably higher HASPI compared with combinations of NAL-R and NR-only systems, namely DNN-SDR\,+\,NAL-R and DNN-NR\,+\,NAL-R.
In fact, DNN-CNRHLC even matches the HASPI obtained by applying NAL-R to the oracle clean speech signal (``Clean\,+\,NAL-R'').
However, systems using NAL-R for HLC show substantially higher HASQI compared with other systems.
This can be explained by the fact that HASQI applies NAL-R to the reference signal internally before it is compared with the processed signal, as illustrated by the near-perfect HASQI achieved by ``Clean\,+\,NAL-R''.
Therefore, HASQI should be interpreted cautiously when evaluating novel HLC algorithms, as it implicitly assumes that NAL-R is the ideal compensation strategy.
Regarding NRMSE, the systems trained for HLC with the proposed differentiable auditory model naturally achieve the best results, while NAL-R shows poor performance similar to the unprocessed noisy signal.

The adjustable system DNN-CNRHLC achieves superior results across all metrics compared with the nonadjustable system DNN-NRHLC.
Thus, the proposed framework not only allows independent adjustment of NR and HLC, but also improves the metrics, while requiring only a slight increase in the number of trainable parameters.
Adjusting~$\alpha_\mathrm{NR}$ and~$\alpha_\mathrm{HLC}$ based on the listener profile or the acoustic scene may further improve the results, as shown by DNN-CNRHLC$^*$, which achieves superior HASPI, HASQI, and NRMSE compared with DNN-CNRHLC.
This further highlights the potential of adaptive parameter selection.
Chaining two DNNs trained for NR and HLC separately (DNN-NR\,+\,DNN-HLC) achieves worse HASPI, HASQI, and NRMSE than DNN-NRHLC and DNN-CNRHLC, while roughly doubling the number of trainable parameters.
These results show the benefit of using a single jointly trained DNN rather than two separately optimized systems applied in series.

\subsection{Paired analysis of main comparisons}
\label{sec:paired}

\begin{table}
\caption{Paired A vs.\ B comparisons. Differences are computed as A~$-$~B over matched scenes and audiograms (NH audiogram only for PESQ, ESTOI, and SDR). 95\% CI is the confidence interval of the mean difference. $d_z$ is the paired Cohen's~$d$. ``A better'' is the percentage of conditions on which A strictly outperforms B.}
\label{tab:paired}
\centering
\setlength{\tabcolsep}{5.5pt}
\begin{tabular}{lS[table-format=-2.3]cS[table-format=-1.2]S[table-format=2.1]}
\toprule
Metric & {Mean $\Delta$} & {95\% CI} & {$d_z$} & {A better (\%)} \\
\midrule
\multicolumn{5}{l}{\textit{DNN-CNRHLC vs.\ DNN-NRHLC}} \\[1pt]
HASPI (\%) $\uparrow$ & 0.80 & [0.71, 0.89] & 0.17 & 60.4 \\
HASQI (\%) $\uparrow$ & 0.95 & [0.91, 0.98] & 0.49 & 70.9 \\
NRMSE (\%) $\downarrow$ & -0.27 & [$-$0.30, $-$0.25] & -0.23 & 65.5 \\
\midrule
\multicolumn{5}{l}{\textit{DNN-CNRHLC vs.\ DNN-NR\,+\,DNN-HLC}} \\[1pt]
HASPI (\%) $\uparrow$ & 2.83 & [2.67, 2.99] & 0.32 & 59.7 \\
HASQI (\%) $\uparrow$ & 3.22 & [3.12, 3.32] & 0.59 & 83.9 \\
NRMSE (\%) $\downarrow$ & -0.78 & [$-$0.82, $-$0.74] & -0.35 & 75.0 \\
\midrule
\multicolumn{5}{l}{\textit{DNN-CNRHLC vs.\ DNN-SDR}} \\[1pt]
PESQ $\uparrow$ & -0.010 & [$-$0.019, $-$0.001] & -0.07 & 41.1 \\
ESTOI (\%) $\uparrow$ & -0.24 & [$-$0.39, $-$0.08] & -0.09 & 44.2 \\
SDR $\uparrow$ & -0.26 & [$-$0.29, $-$0.22] & -0.46 & 21.4 \\
\midrule
\multicolumn{5}{l}{\textit{DNN-CNRHLC$^*$ vs.\ DNN-CNRHLC}} \\[1pt]
HASPI (\%) $\uparrow$ & 3.85 & [3.73, 3.96] & 0.65 & 96.5 \\
HASQI (\%) $\uparrow$ & 2.00 & [1.95, 2.04] & 0.78 & 94.3 \\
NRMSE (\%) $\downarrow$ & -1.56 & [$-$1.59, $-$1.54] & -1.25 & 95.3 \\
\bottomrule
\end{tabular}
\end{table}

\noindent Table~\ref{tab:paired} reports paired comparisons between selected systems.
Each system pair is evaluated on identical scenes and audiograms, yielding \num{11000} paired observations for HASPI, HASQI, and NRMSE, and \num{1000} for PESQ, ESTOI, and SDR.
Differences in HASPI, HASQI, and NRMSE are expressed in percentage points.
Given the large number of observations, the confidence intervals are narrow and exclude zero for all metrics.
Therefore, we focus on the effect size~$d_z$ and the proportion of conditions favoring each system.

The proposed adjustable system DNN-CNRHLC improves on the nonadjustable DNN-NRHLC across all three metrics.
The HASPI difference is small ($d_z = 0.17$), whereas the HASQI difference is moderate and more consistent ($d_z = 0.49$, with DNN-CNRHLC preferred in 70.9\% of conditions).
The flexibility provided by the proposed multi-task framework therefore does not come at the cost of performance, and provides a modest but systematic benefit with only a slight increase in the number of trainable parameters.

The comparison with the cascade DNN-NR\,+\,DNN-HLC shows clear differences in favor of the proposed single-DNN solution.
DNN-CNRHLC achieves higher HASPI ($\Delta = 2.83$, $d_z = 0.32$) and higher HASQI ($\Delta = 3.22$, $d_z = 0.59$), the latter favoring the single DNN in 83.9\% of conditions.
The auditory-model output is also more closely restored ($d_z = -0.35$ on NRMSE, favoring DNN-CNRHLC in 75.0\% of conditions).
Since the cascade uses roughly twice as many trainable parameters, these results suggest that joint training within a single network is more effective than applying two separately trained systems in series.

Compared with the NR-only baseline DNN-SDR, DNN-CNRHLC shows negligible differences on PESQ and ESTOI ($|d_z| < 0.1$) and a larger deficit on SDR ($\Delta = \qty{-0.26}{\decibel}$, $d_z = -0.46$).
The latter is expected, since SDR is the training objective of DNN-SDR, whereas DNN-CNRHLC is optimized to restore the auditory-model output.
The small effect sizes on PESQ and ESTOI indicate that the proposed training objective is competitive with the SDR loss on these two perceptual metrics, at the cost of a small reduction in sample-level fidelity.

Finally, the oracle system DNN-CNRHLC$^*$ improves on DNN-CNRHLC by 3.85 in HASPI ($d_z = 0.65$) and 2.00 in HASQI ($d_z = 0.78$).
Since the oracle selects the exponents~$\alpha_\mathrm{NR}$ and~$\alpha_\mathrm{HLC}$ from a grid that contains the fixed pair used by DNN-CNRHLC, it always matches or outperforms it.
The reported percentages thus indicate the proportion of conditions in which DNN-CNRHLC does not use the optimal exponent pair for that metric.
This proportion exceeds 94\% for all three metrics, which shows that the optimal balance between NR and HLC is strongly condition-dependent, and highlights the potential of a strategy for adapting~$\alpha_\mathrm{NR}$ and~$\alpha_\mathrm{HLC}$ to the listener and the acoustic scene.

\subsection{Example spectrograms and masks}
\label{sec:results_example}

\begin{figure*}
  \centering
  \includegraphics[width=.99\linewidth]{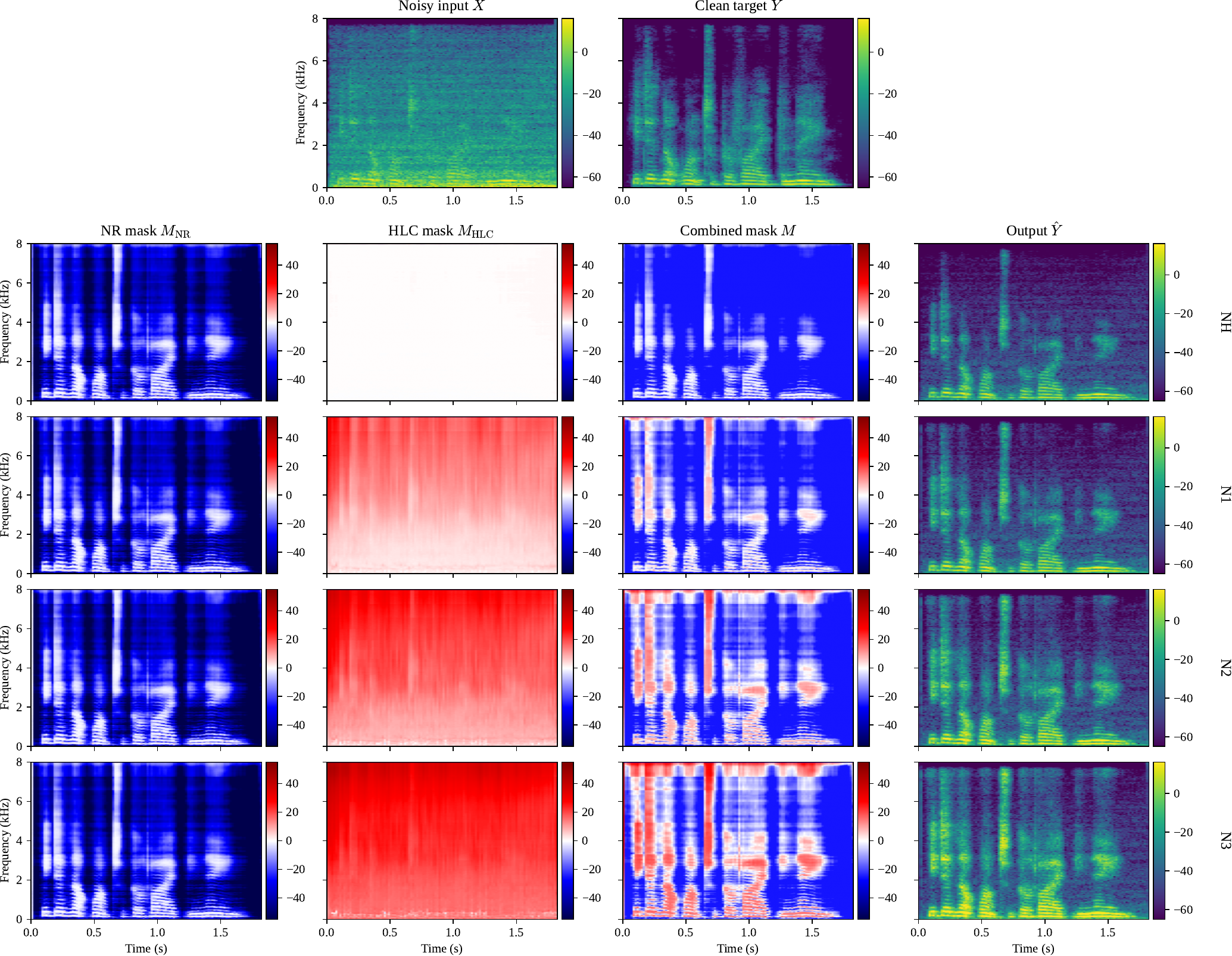}
  \caption{
    Example input spectrogram~$X$, clean spectrogram~$Y$, predicted NR mask~$M_\mathrm{NR}$, HLC mask~$M_\mathrm{HLC}$, combined mask~$M$, and output spectrogram~$\hat{Y}$ for the NH audiogram and HI audiograms N1, N2, and N3.
    Both~$\alpha_\mathrm{NR}$ and~$\alpha_\mathrm{HLC}$ are set to 1.
    The masks are complex-valued and their magnitude is plotted in dB.
    The magnitude of the combined mask~$M$ is thresholded using a minimum gain~$G_{\min}$ of \qty{-25}{\decibel}.
  }
  \label{fig:example_mask_combination}
\end{figure*}

\noindent Figure~\ref{fig:example_mask_combination} shows an example of a noisy input spectrogram and a clean target spectrogram, the NR and HLC masks~$M_\mathrm{NR}$ and~$M_\mathrm{HLC}$ predicted by DNN-CNRHLC, the resulting combined mask~$M$, and the output spectrogram~$\hat{Y}$ for listener profiles NH, N1, N2, and N3.
For illustration purposes, both~$\alpha_\mathrm{NR}$ and~$\alpha_\mathrm{HLC}$ are set to 1, resulting in maximum contribution of the predicted NR and HLC masks.
The NR mask~$M_\mathrm{NR}$ attenuates TF units dominated by noise and largely preserves those with speech activity.
It is also consistent across listener profiles.
Note that this desirable consistency is not explicitly enforced but emerges from the proposed optimization procedure.
Meanwhile, the magnitude of the HLC mask~$M_\mathrm{HLC}$ is \qty{0}{\decibel} for the NH audiogram, which is desirable since no HLC is required.
It increases as the severity of the hearing loss increases, particularly at high frequencies where the hearing loss is more pronounced.
Note that while~$M_\mathrm{HLC}$ shows fewer temporal fluctuations than~$M_\mathrm{NR}$, it is not constant over time.
This indicates that the learned compensation is a dynamic rather than a purely static frequency-dependent amplification.

The combined mask~$M$ effectively integrates both NR and HLC, as it correctly attenuates TF units dominated by noise, and amplifies those dominated by speech.
This is because, for TF units dominated by noise, the attenuation provided by~$M_\mathrm{NR}$ is much greater than the amplification provided by~$M_\mathrm{HLC}$ in absolute value.
In that regard, it is important to threshold the magnitude of the combined mask~$M$ with~$G_{\min}$ only after the mask combination, as thresholding~$M_\mathrm{NR}$ before combining with~$M_\mathrm{HLC}$ would result in TF units dominated by noise being amplified.
This illustrates the benefit of designing the NR and HLC stages together, since information about the noise presence can guide the HLC strategy.

\subsection{Influence of center frequencies on long-term average gain}
\label{sec:results_ltag}

\begin{figure}
    \centering
    \includegraphics[scale=0.68]{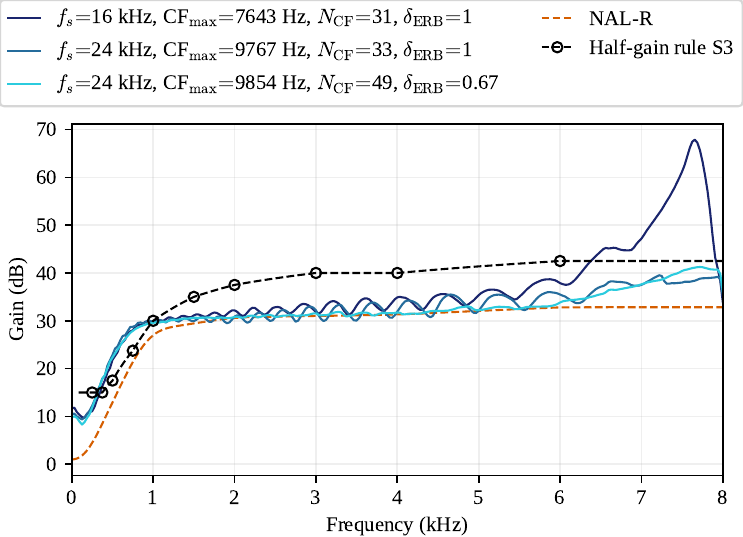}
    \caption{Long-term average gain prescribed by DNN-CNRHLC (blue, $\alpha_\mathrm{NR} = 0$, $\alpha_\mathrm{HLC} = 1$) and NAL-R (orange) for audiogram S3. Shades of blue indicate different auditory-model filterbank configurations. The half-gain rule (black) is half the audiogram thresholds interpolated to the filterbank center frequencies.}
    \label{fig:ltag}
\end{figure}

\noindent Similar to~\cite{leer2025hearing}, we compute the long-term average gain prescribed by the proposed DNN-CNRHLC as a function of frequency using Welch's method~\cite{welch1967use}.
We use stationary white noise at \qty{70}{\decibel} SPL as the input signal, average the power spectral densities over 100 independent 4-\unit{\second}-long noise realizations, fix~$\alpha_\mathrm{NR} = 0$ and~$\alpha_\mathrm{HLC} = 1$, and compare with NAL-R for the audiogram S3 in Fig.~\ref{fig:ltag}.
During initial experiments, we set the operating sampling rate of the auditory model to~$f_s=\qty{16}{\kilo\hertz}$, matching the input speech signals, and used a center-frequency spacing of~$\delta_\mathrm{ERB}=1$ up to~$\mathrm{CF}_\mathrm{max} = \qty{7643}{\hertz}$ (dark blue curve).
However, this resulted in a ripple-like structure as in~\cite{leer2025hearing}, and excessive amplification in the \qtyrange{6}{8}{\kilo\hertz} range corresponding to the two to three highest center frequencies.
We found that the latter was due to the absence of auditory filters above the frequency range of the input signal: in contrast to middle frequencies, amplifying high-frequency content only excited lower center frequencies.
Since there were no auditory filters above Nyquist whose response should be restored to that of the NH auditory model, the DNN restored the response of the two to three highest center frequencies without being penalized by higher center frequencies, which resulted in greater gain compared with middle frequencies.
We addressed this by increasing the operating sampling rate of the auditory model to~$f_s=\qty{24}{\kilo\hertz}$, such that center frequencies could be added up to~$\mathrm{CF}_\mathrm{max} = \qty{9767}{\hertz}$ (medium blue curve).
This considerably reduced the excessive amplification in the \qtyrange{6}{8}{\kilo\hertz} range, but preserved the ripple-like structure.
Following~\cite{leer2025hearing}, we increased the number of center frequencies by reducing their spacing.
More specifically, we set~$\delta_\mathrm{ERB}=0.67$, which was effective at smoothing the long-term average gain (light blue curve).
This filterbank configuration is the one used for training all DNN-based HLC systems.

Compared with NAL-R, the proposed DNN-CNRHLC prescribes a very similar long-term average gain in the \qtyrange{1}{6}{\kilo\hertz} range (maximum deviation of \qty{2.6}{\decibel}).
In the \qtyrange{0}{1}{\kilo\hertz} range, DNN-CNRHLC prescribes \qty{7.2}{\decibel} more gain than NAL-R on average, with a maximum difference of \qty{9.6}{\decibel}.
In the \qtyrange{6}{8}{\kilo\hertz} range, it prescribes \qty{4.8}{\decibel} more gain on average, and the maximum difference is \qty{8.4}{\decibel}.
More generally, the gain provided by DNN-CNRHLC is broadly comparable to NAL-R, especially in the range contributing most to speech intelligibility.
The HASPI benefit of the proposed system over NAL-R in Table~\ref{tab:results} therefore does not appear to be explained by greater average amplification in this frequency range.

In contrast to NAL-R, which prescribes a linear and level-independent gain, the proposed system inherits the level dependence of the auditory model, such that the prescribed gain varies with the input level.
We verified that the other standard audiograms and input levels between \qtylist{65;85}{\decibel} SPL yield qualitatively similar curves.
Overall, the gains learned by the proposed system are comparable to an established prescription rule under the evaluated conditions.
We stress that this characterizes the gain prescribed by the system, and does not constitute an output-limiting mechanism; ensuring the output remains below uncomfortable levels requires appropriate explicit constraints and output-level verification.

\subsection{Gain distribution for noisy speech}
\label{sec:results_gain_dist}

\begin{figure}
  \centering
  \includegraphics[scale=0.68]{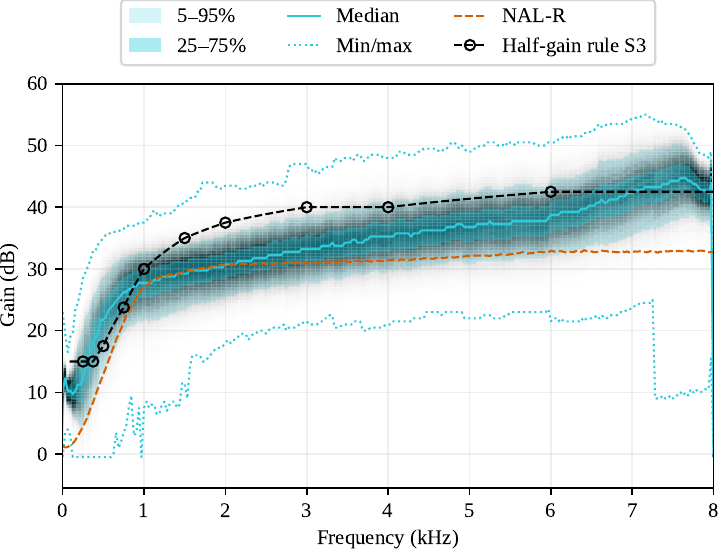}
  \caption{
    Distribution of the gain applied by DNN-CNRHLC ($\alpha_\mathrm{NR} = 0$, $\alpha_\mathrm{HLC} = 1$) in each TF unit for 300 noisy test scenes and audiogram S3.
    The gray shading shows the distribution at each frequency, with darker shades indicating more TF units.
    The blue line and shaded areas show the median and the \qtyrange{25}{75}{\percent} and \qtyrange{5}{95}{\percent} ranges, and the dotted lines show the minimum and maximum.
    TF units more than \qty{40}{\decibel} below the maximum input level in the scene are excluded.
    The gain prescribed by NAL-R (orange) and the half-gain rule (black) are shown for reference.
  }
  \label{fig:gain_dist}
\end{figure}

\begin{figure}
  \centering
  \includegraphics[scale=0.68]{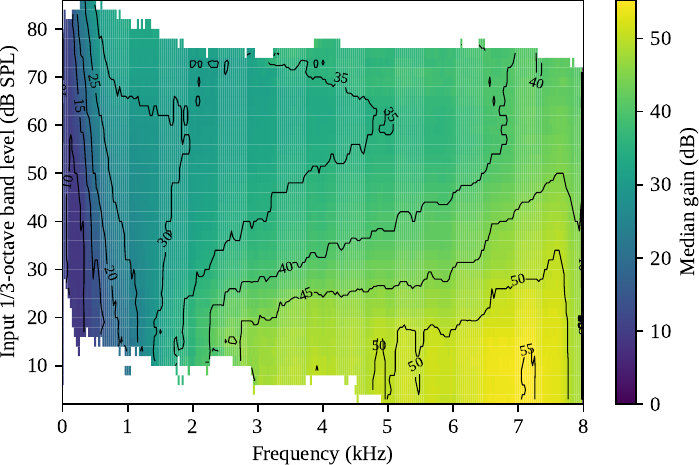}
  \caption{
    Median gain applied by DNN-CNRHLC ($\alpha_\mathrm{NR} = 0$, $\alpha_\mathrm{HLC} = 1$) as a function of frequency and input level for the same scenes and audiogram as in Fig.~\ref{fig:gain_dist}.
    The input level of each TF unit is the level of the noisy input in a one-third-octave band centered at that frequency.
    Contour lines are drawn every \qty{5}{\decibel}.
    White areas correspond to combinations of frequency and level with fewer than 50 TF units.
  }
  \label{fig:gain_level_map}
\end{figure}

\noindent The long-term average gain in Section~\ref{sec:results_ltag} is obtained with a stationary input at a single level, and thus only provides a partial view of the gain applied by a nonlinear system to speech in noise.
To characterize gain variability under realistic conditions, we compute the magnitude of the final mask~$|M|$ in dB for each TF unit of 300 noisy and reverberant test scenes, whose levels range from \qtyrange{65}{85}{\decibel} SPL and SNRs from \qtyrange{-5}{15}{\decibel}.
As in Section~\ref{sec:results_ltag}, we use audiogram S3, and set~$\alpha_\mathrm{NR} = 0$ and~$\alpha_\mathrm{HLC} = 1$ to isolate HLC.
To focus on TF units contributing to the output, we exclude those that are more than \qty{40}{\decibel} below the maximum input level in the scene.
The resulting distribution is shown in Fig.~\ref{fig:gain_dist}, together with the gain prescribed by NAL-R and the half-gain rule, i.e.\ half of the audiogram thresholds.

Below \qty{1}{\kilo\hertz}, the median gain roughly follows the half-gain rule, with an average absolute difference of \qty{2.8}{\decibel}.
Meanwhile, it exceeds NAL-R by \qty{6.9}{\decibel} on average.
In the \qtyrange{1}{6}{\kilo\hertz} range, it lies between the two prescriptions, exceeding NAL-R by \qty{2.7}{\decibel} on average and remaining \qty{5.4}{\decibel} below the half-gain rule on average.
Above \qty{6}{\kilo\hertz}, it rises toward the half-gain rule, with an average absolute difference of \qty{1.6}{\decibel}, and exceeds NAL-R by \qty{9.1}{\decibel} on average.
Compared with the long-term average gain for white noise in Fig.~\ref{fig:ltag}, the median gain is higher at high frequencies, which is consistent with the lower high-frequency levels of speech and the level dependence discussed below.
The gain also varies considerably across TF units: the interquartile range is about \qty{5}{\decibel} and the \qtyrange{5}{95}{\percent} range about \qtyrange{11}{13}{\decibel} at all frequencies.
This confirms that DNN-CNRHLC applies a dynamic gain rather than a static frequency-dependent amplification, consistent with the HLC mask shown in Fig.~\ref{fig:example_mask_combination}.
Nevertheless, the gain remains within a moderate range around the median, and the extreme values are reached by only a small fraction of TF units.

To analyze the level dependence, Fig.~\ref{fig:gain_level_map} shows the median gain as a function of frequency and input level.
Above \qty{2}{\kilo\hertz}, the gain decreases as the input level increases, consistent with compressive amplification.
For example, at \qty{4}{\kilo\hertz}, the median gain decreases from \qty{40.8}{\decibel} at an input level of \qty{30}{\decibel} SPL to \qty{35.3}{\decibel} at \qty{50}{\decibel} SPL, and remains roughly constant above.
Between \qtylist{30;70}{\decibel} SPL, the gain decreases by about \qty{0.2}{\decibel} per dB of input level in the \qtyrange{4}{7}{\kilo\hertz} range, which corresponds to a compression ratio of about 1.25.
Around \qty{2}{\kilo\hertz}, the gain is approximately independent of the input level.
Below \qty{1}{\kilo\hertz}, the gain increases with the input level, e.g.\ from \qty{14.8}{\decibel} to \qty{26.3}{\decibel} at \qty{500}{\hertz} between \qtylist{30;70}{\decibel} SPL.

\section{Discussion}
\label{sec:discussion}

\subsection{Optimizing the balance between NR and HLC}

\noindent The results for the proposed DNN-CNRHLC were obtained after fixing~$\alpha_\mathrm{NR}$ and~$\alpha_\mathrm{HLC}$ across all test scenes and listener profiles.
However, this choice may be suboptimal, since the appropriate balance between NR and HLC may depend on the acoustic properties of the scene, such as the SNR, the type of background noise, or the reverberation time.
It may also depend on the listener profile, the listening intent, or individual preference.
This is supported by the results for DNN-CNRHLC$^*$, which show that selecting optimal~$\alpha_\mathrm{NR}$ and~$\alpha_\mathrm{HLC}$ for each listener, scene, and metric can further improve the results compared with using fixed values across all configurations.
While this highlights the potential of the proposed framework, future work should investigate strategies for optimizing~$\alpha_\mathrm{NR}$ and~$\alpha_\mathrm{HLC}$ on a per-listener and per-scene basis.
For example, since the mask exponentiation and combination are computationally lightweight and do not introduce additional algorithmic latency, $\alpha_\mathrm{NR}$ and~$\alpha_\mathrm{HLC}$ could be adjusted on-the-fly based on short-time acoustic features.
Alternatively, a user interface could allow listeners to adjust~$\alpha_\mathrm{NR}$ and~$\alpha_\mathrm{HLC}$ manually in real time according to their own preferences, which may reveal insights into subjectively optimal strategies for joint NR and HLC.

\subsection{Validating the differentiable auditory model}

\noindent As discussed in Section~\ref{sec:setup_am}, the differentiable auditory model used in this study is a simplified auditory front-end inspired by established models of auditory processing and masking~\cite{dau1996quantitative,dau1997modeling,jepsen2008computational,paulick2025computational}.
Auditory models within this class have been shown to account for a wide range of psychoacoustic phenomena in NH listeners, including intensity discrimination, spectral and temporal masking, and modulation detection~\cite{jepsen2008computational}.
Related model extensions have also been used to describe the average behavior of HI listeners~\cite{jepsen2011characterizing,relano2022speech}.
However, the simplifications introduced in the present work may affect the extent to which the resulting model retains these predictive properties.
Future work should therefore validate the proposed differentiable auditory model, for example by examining whether its output supports accurate predictions of NH and HI listener behavior across relevant psychoacoustic conditions.
Such analyses may reveal limitations of the current model formulation and thereby suggest ways of improving both the auditory model itself and the resulting downstream NR and HLC performance of the DNN.

\subsection{Conflicting NR and HLC objectives}

\begin{figure}
    \centering
    \includegraphics[width=0.85\linewidth]{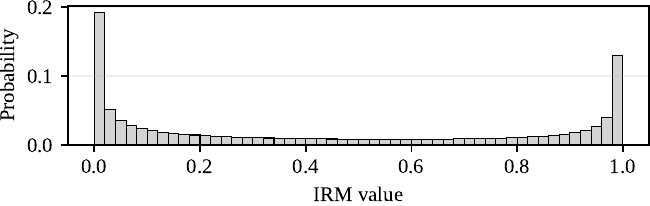}
    \caption{Distribution of IRM values across all TF units in the test dataset.}
    \label{fig:irm_distribution}
\end{figure}

\noindent NR and HLC inherently pursue conflicting objectives: NR aims to attenuate noise whereas HLC aims to restore audibility through level-dependent amplification.
While this conflict is negligible in TF units dominated by either speech or noise, it may become relevant for TF units in which speech and noise exhibit comparable energy.
Importantly, this challenge is not unique to the proposed framework but is common to any approach combining NR and HLC, including traditional signal processing chains in hearing aids.
In practice, however, the impact of this conflict may be limited, since most TF units are expected to be dominated by either speech or noise.
This assumption underlies established speech enhancement approaches such as ideal binary masking and computational auditory scene analysis~\cite{wang2005ideal,wang2006computational}, scene-aware dynamic range compression strategies for hearing aids~\cite{hassager2017preserving,may2018signal,may2020scene}, and a recent test-time adaptation method for mask-based speech enhancement~\cite{raichle2026test}.
To examine this assumption, we analyze the distribution of ideal ratio mask (IRM)~\cite{wang2014training} values across all TF units in the test dataset.
As shown in Fig.~\ref{fig:irm_distribution}, the distribution is strongly bimodal, with most IRM values clustering near zero (strongly noise-dominated) or one (strongly speech-dominated).
This indicates that TF units with strongly conflicting NR and HLC objectives are relatively rare.
Finally, since the predicted masks are complex-valued, the proposed system can, in principle, recover and amplify the clean speech even in TF units where both speech and noise are prominent, by modifying both the magnitude and the phase.

\subsection{Scope and clinical validation}

\noindent The present study focuses on the development and objective evaluation of the proposed adjustable joint NR and HLC framework.
Several limitations therefore need to be addressed before the framework can be considered for clinical applications.
First, although the DNN was trained with independently perturbed audiometric thresholds, the underlying profiles are limited to flat and high-frequency-sloping audiograms.
The present results therefore do not establish performance for other audiometric configurations, such as rising, cookie-bite, or highly atypical hearing losses.
Evaluating a broader range of clinically observed audiograms will be important in future work.

Second, the improvements reported here are based on objective metrics and auditory-model responses.
They should not be interpreted as direct evidence of improved speech intelligibility, sound quality, or listening comfort in HI listeners.
Behavioral validation with HI participants is therefore essential for determining whether the objective benefits observed here translate into perceptual benefit.

Third, the present framework addresses the computational optimization of NR and HLC rather than a complete clinical hearing-aid fitting procedure.
Practical implementation would additionally require individualized electroacoustic calibration and real-ear verification, as well as appropriate output-level constraints.
The gain analyses in Sections~\ref{sec:results_ltag} and~\ref{sec:results_gain_dist} characterize the behavior of the learned HLC under the evaluated conditions, but do not provide a deterministic guarantee of safe output levels.
Although an explicit maximum-gain constraint can be applied to the predicted mask through~$G_{\max}$, clinical implementation would require appropriate output limiting and verification based on individual loudness tolerance.

Finally, the current DNN architecture is noncausal and has an algorithmic latency that is too high for direct real-time hearing-aid implementation.
Adapting the framework to causal, low-latency processing therefore represents another important step toward practical deployment.

\section{Conclusion}
\label{sec:conclusion}

\noindent In this work, we proposed a multi-task learning framework for optimizing a DNN to perform joint NR and HLC.
In contrast to recent approaches, we used a differentiable auditory model, which eliminated the need to train an auditory-model emulator and therefore enabled end-to-end optimization.
This simplification of the optimization framework facilitates refining the auditory model to improve the DNN performance.
In addition, the listener's audiogram was provided as an extra input to the DNN, enabling personalization without retraining, unlike in many previous studies.
To the best of our knowledge, this is the first study to use an auditory model to train a DNN for both NR and HLC across a wide range of flat and high-frequency-sloping profiles.

Furthermore, the proposed solution allows adjusting NR and HLC independently at inference time, without retraining the DNN.
This was achieved by defining two distinct training objectives, one for NR and one for HLC, and by predicting a separate TF mask for each task.
During training, the contributions of the two objectives were automatically balanced using an uncertainty-based weighting scheme~\cite{kendall2018multi}.
After training, the two masks can be exponentiated independently and then combined, allowing flexible control over NR and HLC.
This represents a significant improvement over previous approaches in which both tasks were optimized through a single training objective, resulting in systems that deliver a fixed combination of NR and HLC.
In addition to this increased flexibility, the proposed approach also showed superior joint NR and HLC performance according to objective metrics once the exponents~$\alpha_\mathrm{NR}$ and~$\alpha_\mathrm{HLC}$ were tuned.
The proposed system also outperformed NAL-R in terms of HASPI and restoration of the auditory-model output.
Although NAL-R achieved higher HASQI, this is likely due to the fact that HASQI implicitly assumes that NAL-R is the ideal compensation strategy.
The proposed system also matched a specialized NR-only system on PESQ and ESTOI despite additionally performing HLC, which may be explained by the multi-task learning framework: by learning internal representations that are useful for both tasks, the DNN may become more robust~\cite{baxter2000model}.
Finally, applying two specialized systems in series for joint NR and HLC, similar to conventional hearing-aid processing chains, resulted in inferior performance compared with the proposed single-DNN system, despite doubling the total number of trainable parameters.

The flexibility offered by the proposed system opens several avenues for future work.
The exponents~$\alpha_\mathrm{NR}$ and~$\alpha_\mathrm{HLC}$ could be adjusted heuristically on the basis of short-time acoustic features to further improve performance and perceptual benefit.
Alternatively, future experiments could allow participants to set~$\alpha_\mathrm{NR}$ and~$\alpha_\mathrm{HLC}$ according to personal preference, which may provide insight into subjectively optimal strategies for joint NR and HLC.
Future work should also focus on validating the proposed differentiable auditory model and on adapting the DNN architecture to reduce computational complexity and enable real-time processing.
Finally, the present study was limited to objective metrics; future work should therefore assess the perceptual benefit of the proposed solution through subjective listening tests involving HI participants.

\bibliographystyle{IEEEtran}
\bibliography{bib/IEEEabrv, bib/abbrv, bib/bib}

\begin{thebibliography}{100}
\providecommand{\url}[1]{#1}
\csname url@samestyle\endcsname
\providecommand{\newblock}{\relax}
\providecommand{\bibinfo}[2]{#2}
\providecommand{\BIBentrySTDinterwordspacing}{\spaceskip=0pt\relax}
\providecommand{\BIBentryALTinterwordstretchfactor}{4}
\providecommand{\BIBentryALTinterwordspacing}{\spaceskip=\fontdimen2\font plus
\BIBentryALTinterwordstretchfactor\fontdimen3\font minus \fontdimen4\font\relax}
\providecommand{\BIBforeignlanguage}[2]{{%
\expandafter\ifx\csname l@#1\endcsname\relax
\typeout{** WARNING: IEEEtran.bst: No hyphenation pattern has been}%
\typeout{** loaded for the language `#1'. Using the pattern for}%
\typeout{** the default language instead.}%
\else
\language=\csname l@#1\endcsname
\fi
#2}}
\providecommand{\BIBdecl}{\relax}
\BIBdecl

\bibitem{world2021world}
\BIBentryALTinterwordspacing
{World Health Organization}, ``World report on hearing,'' 2021. [Online]. Available: \url{https://www.who.int/publications/i/item/world-report-on-hearing}
\BIBentrySTDinterwordspacing

\bibitem{georganti2020intelligent}
E.~Georganti, G.~Courtois, P.~Derleth, and S.~Launer, ``Intelligent hearing instruments---{Trends} and challenges,'' in \emph{The Technology of Binaural Understanding}.\hskip 1em plus 0.5em minus 0.4em\relax Springer, 2020, pp. 733--761.

\bibitem{wang2018supervised}
D.~Wang and J.~Chen, ``Supervised speech separation based on deep learning: {An} overview,'' \emph{{IEEE/ACM} Trans. Audio, Speech, Lang. Process.}, vol.~26, pp. 1702--1726, 2018.

\bibitem{zheng2023sixty}
C.~Zheng, H.~Zhang, W.~Liu, X.~Luo, A.~Li, X.~Li, and B.~C.~J. Moore, ``Sixty years of frequency-domain monaural speech enhancement: {From} traditional to deep learning methods,'' \emph{Trends Hear.}, vol.~27, p. 23312165231209913, 2023.

\bibitem{graetzer2021clarity}
S.~Graetzer, J.~Barker, T.~J. Cox, M.~A. Akeroyd, J.~F. Culling, G.~Naylor, E.~Porter, and R.~Viveros~Mu{\~n}oz, ``Clarity-2021 challenges: {Machine} learning challenges for advancing hearing aid processing,'' in \emph{Proc. Interspeech}, 2021, pp. 686--690.

\bibitem{baby2021convolutional}
D.~Baby, A.~van~den Broucke, and S.~Verhulst, ``A convolutional neural-network model of human cochlear mechanics and filter tuning for real-time applications,'' \emph{Nat. Mach. Intell.}, vol.~3, pp. 134--143, 2021.

\bibitem{drakopoulos2021convolutional}
F.~Drakopoulos, D.~Baby, and S.~Verhulst, ``A convolutional neural-network framework for modelling auditory sensory cells and synapses,'' \emph{Commun. Biol.}, vol.~4, p. 827, 2021.

\bibitem{nagathil2021computationally}
A.~Nagathil, F.~G{\"o}bel, A.~Nelus, and I.~C. Bruce, ``Computationally efficient {DNN}-based approximation of an auditory model for applications in speech processing,'' in \emph{Proc. ICASSP}, 2021, pp. 301--305.

\bibitem{nagathil2023wavenet}
A.~Nagathil and I.~C. Bruce, ``{WaveNet}-based approximation of a cochlear filtering and hair cell transduction model,'' \emph{J. Acoust. Soc. Am.}, vol. 154, pp. 191--202, 2023.

\bibitem{leer2024train}
P.~Leer, J.~Jensen, Z.-H. Tan, J.~{\O}stergaard, and L.~Bramsl{\o}w, ``How to train your ears: {Auditory}-model emulation for large-dynamic-range inputs and mild-to-severe hearing losses,'' \emph{{IEEE/ACM} Trans. Audio, Speech, Lang. Process.}, vol.~32, pp. 2006--2020, 2024.

\bibitem{drakopoulos2022differentiable}
F.~Drakopoulos and S.~Verhulst, ``A differentiable optimisation framework for the design of individualised {DNN}-based hearing-aid strategies,'' in \emph{Proc. ICASSP}, 2022, pp. 351--355.

\bibitem{drakopoulos2023neural}
------, ``A neural-network framework for the design of individualised hearing-loss compensation,'' \emph{{IEEE/ACM} Trans. Audio, Speech, Lang. Process.}, vol.~31, pp. 2395--2409, 2023.

\bibitem{drakopoulos2023dnn}
F.~Drakopoulos, A.~van~den Broucke, and S.~Verhulst, ``A {DNN}-based hearing-aid strategy for real-time processing: {One} size fits all,'' in \emph{Proc. ICASSP}, 2023, pp. 1--5.

\bibitem{wen2025dconnear}
C.~Wen, G.~Torfs, and S.~Verhulst, ``{dCoNNear}: {An} artifact-free neural network architecture for closed-loop audio signal processing,'' \emph{IEEE Trans. Audio, Speech, Lang. Process.}, vol.~33, pp. 4414--4429, 2025.

\bibitem{wen2025individualized}
C.~Wen and S.~Verhulst, ``Individualized speech enhancement for hearing-impaired listeners,'' in \emph{Proc. Interspeech}, 2025, pp. 3843--3847.

\bibitem{leer2025hearing}
P.~Leer, J.~Jensen, L.~H. Carney, Z.-H. Tan, J.~{\O}stergaard, and L.~Bramsl{\o}w, ``Hearing-loss compensation using deep neural networks: {A} framework and results from a listening test,'' \emph{IEEE Trans. Audio, Speech, Lang. Process.}, vol.~33, pp. 828--841, 2025.

\bibitem{tu2021dhasp}
Z.~Tu, N.~Ma, and J.~Barker, ``{DHASP}: {Differentiable} hearing aid speech processing,'' in \emph{Proc. ICASSP}, 2021, pp. 296--300.

\bibitem{tu2021optimising}
------, ``Optimising hearing aid fittings for speech in noise with a differentiable hearing loss model,'' in \emph{Proc. Interspeech}, 2021, pp. 691--695.

\bibitem{tu2021two}
Z.~Tu, J.~Zhang, N.~Ma, and J.~Barker, ``A two-stage end-to-end system for speech-in-noise hearing aid processing,'' in \emph{Proc. Clarity}, 2021, pp. 19--23.

\bibitem{drgas2024dynamic}
S.~Drgas, L.~Bramsl{\o}w, A.~Politis, G.~Naithani, and T.~Virtanen, ``Dynamic processing neural network architecture for hearing loss compensation,'' \emph{{IEEE/ACM} Trans. Audio, Speech, Lang. Process.}, vol.~32, pp. 203--214, 2024.

\bibitem{zmolikova2021but}
K.~\v{Z}mol\'{i}kov\'{a} and J.~\v{C}ernock\'y, ``{BUT} system for the first {Clarity} {Enhancement} {Challenge},'' in \emph{Proc. Clarity}, 2021, pp. 5--7.

\bibitem{cheng2023speech}
J.~Cheng, R.~Liang, L.~Zhao, C.~Huang, and B.~W. Schuller, ``Speech denoising and compensation for hearing aids using an {FTCRN}-based metric {GAN},'' \emph{{IEEE} Signal Process. Lett.}, vol.~30, pp. 374--378, 2023.

\bibitem{ni2026affine}
Y.~Ni, R.~Liang, X.~Hao, J.~Cheng, Q.~Wang, C.~Huang, C.~Zou, W.~Zhou, W.~Ding, and B.~W. Schuller, ``Affine modulation-based audiogram fusion network for joint noise reduction and hearing loss compensation,'' \emph{Inf. Fusion}, vol. 127, p. 103726, 2026.

\bibitem{baer1993effects}
T.~Baer and B.~C.~J. Moore, ``Effects of spectral smearing on the intelligibility of sentences in noise,'' \emph{J. Acoust. Soc. Am.}, vol.~94, pp. 1229--1241, 1993.

\bibitem{baer1994effects}
------, ``Effects of spectral smearing on the intelligibility of sentences in the presence of interfering speech,'' \emph{J. Acoust. Soc. Am.}, vol.~95, pp. 2277--2280, 1994.

\bibitem{moore1993simulation}
B.~C.~J. Moore and B.~R. Glasberg, ``Simulation of the effects of loudness recruitment and threshold elevation on the intelligibility of speech in quiet and in a background of speech,'' \emph{J. Acoust. Soc. Am.}, vol.~94, pp. 2050--2062, 1993.

\bibitem{stone1999tolerable}
M.~A. Stone and B.~C.~J. Moore, ``Tolerable hearing aid delays. {I}. {Estimation} of limits imposed by the auditory path alone using simulated hearing losses,'' \emph{Ear Hear.}, vol.~20, pp. 182--192, 1999.

\bibitem{baxter2000model}
J.~Baxter, ``A model of inductive bias learning,'' \emph{J. Artif. Intell. Res.}, vol.~12, pp. 149--198, 2000.

\bibitem{killion1993types}
M.~C. Killion, ``The 3 types of sensorineural hearing loss: {Loudness} and intelligibility considerations,'' \emph{Hear. J.}, vol.~46, pp. 31--36, 1993.

\bibitem{neher2014relating}
T.~Neher, ``Relating hearing loss and executive functions to hearing aid users' preference for, and speech recognition with, different combinations of binaural noise reduction and microphone directionality,'' \emph{Front. Neurosci.}, vol.~8, p. 391, 2014.

\bibitem{neher2016directional}
T.~Neher, K.~C. Wagener, and R.-L. Fischer, ``Directional processing and noise reduction in hearing aids: {Individual} and situational influences on preferred setting,'' \emph{J. Am. Acad. Audiol.}, vol.~27, pp. 628--646, 2016.

\bibitem{kendall2018multi}
A.~Kendall, Y.~Gal, and R.~Cipolla, ``Multi-task learning using uncertainty to weigh losses for scene geometry and semantics,'' in \emph{Proc. CVPR}, 2018, pp. 7482--7491.

\bibitem{gonzalez2025controllable}
P.~Gonzalez, T.~Dau, and T.~May, ``Controllable joint noise reduction and hearing loss compensation using a differentiable auditory model,'' in \emph{Proc. Clarity}, 2025, pp. 48--52.

\bibitem{zhao2018perceptually}
Y.~Zhao, B.~Xu, R.~Giri, and T.~Zhang, ``Perceptually guided speech enhancement using deep neural networks,'' in \emph{Proc. ICASSP}, 2018, pp. 5074--5078.

\bibitem{vuong2021modulation}
T.~Vuong, Y.~Xia, and R.~M. Stern, ``A modulation-domain loss for neural-network-based real-time speech enhancement,'' in \emph{Proc. ICASSP}, 2021, pp. 6643--6647.

\bibitem{eng2022using}
N.~Eng, Y.~Hioka, and C.~I. Watson, ``Using perceptual quality features in the design of the loss function for speech enhancement,'' in \emph{Proc. APSIPA ASC}, 2022, pp. 1904--1909.

\bibitem{monir2025frequency}
N.-E. Monir, P.~Magron, and R.~Serizel, ``Frequency-weighted training losses for phoneme-level {DNN}-based speech enhancement,'' in \emph{Proc. MMSP}, 2025, pp. 310--315.

\bibitem{brons2012perceptual}
I.~Brons, R.~Houben, and W.~A. Dreschler, ``Perceptual effects of noise reduction by time-frequency masking of noisy speech,'' \emph{J. Acoust. Soc. Am.}, vol. 132, pp. 2690--2699, 2012.

\bibitem{borgstrom2021speech}
B.~J. Borgstr{\"o}m and M.~S. Brandstein, ``Speech enhancement via attention masking network ({SEAMNET}): {An} end-to-end system for joint suppression of noise and reverberation,'' \emph{{IEEE/ACM} Trans. Audio, Speech, Lang. Process.}, vol.~29, pp. 515--526, 2021.

\bibitem{johnson2024ideal}
E.~M. Johnson and E.~W. Healy, ``An ideal compressed mask for increasing speech intelligibility without sacrificing environmental sound recognition,'' \emph{J. Acoust. Soc. Am.}, vol. 156, pp. 3958--3969, 2024.

\bibitem{cao2022cmgan}
R.~Cao, S.~Abdulatif, and B.~Yang, ``{CMGAN}: {Conformer}-based metric {GAN} for speech enhancement,'' in \emph{Proc. Interspeech}, 2022, pp. 936--940.

\bibitem{abdulatif2024cmgan}
S.~Abdulatif, R.~Cao, and B.~Yang, ``{CMGAN}: {Conformer}-based metric-{GAN} for monaural speech enhancement,'' \emph{{IEEE/ACM} Trans. Audio, Speech, Lang. Process.}, vol.~32, pp. 2477--2493, 2024.

\bibitem{li2022glance}
A.~Li, C.~Zheng, L.~Zhang, and X.~Li, ``Glance and gaze: {A} collaborative learning framework for single-channel speech enhancement,'' \emph{Appl. Acoust.}, vol. 187, p. 108499, 2022.

\bibitem{zhao2023d2former}
S.~Zhao and B.~Ma, ``{D2Former}: {A} fully complex dual-path dual-decoder conformer network using joint complex masking and complex spectral mapping for monaural speech enhancement,'' in \emph{Proc. ICASSP}, 2023, pp. 1--5.

\bibitem{li2021two}
A.~Li, W.~Liu, C.~Zheng, C.~Fan, and X.~Li, ``Two heads are better than one: {A} two-stage complex spectral mapping approach for monaural speech enhancement,'' \emph{{IEEE/ACM} Trans. Audio, Speech, Lang. Process.}, vol.~29, pp. 1829--1843, 2021.

\bibitem{lu2023mp}
Y.-X. Lu, Y.~Ai, and Z.-H. Ling, ``{MP-SENet}: {A} speech enhancement model with parallel denoising of magnitude and phase spectra,'' in \emph{Proc. Interspeech}, 2023, pp. 3834--3838.

\bibitem{lu2025explicit}
------, ``Explicit estimation of magnitude and phase spectra in parallel for high-quality speech enhancement,'' \emph{Neural Netw.}, vol. 189, p. 107562, 2025.

\bibitem{xu2015multi}
Y.~Xu, J.~Du, Z.~Huang, L.-R. Dai, and C.-H. Lee, ``Multi-objective learning and mask-based post-processing for deep neural network based speech enhancement,'' in \emph{Proc. Interspeech}, 2015, pp. 1508--1512.

\bibitem{sun2017multiple}
L.~Sun, J.~Du, L.-R. Dai, and C.-H. Lee, ``Multiple-target deep learning for {LSTM-RNN} based speech enhancement,'' in \emph{Proc. HSCMA}, 2017, pp. 136--140.

\bibitem{zhou2021masks}
L.~Zhou, W.~Jiang, J.~Xu, F.~Wen, and P.~Liu, ``Masks fusion with multi-target learning for speech enhancement,'' arXiv:2109.11164, 2021.

\bibitem{shi2020spectrograms}
H.~Shi, L.~Wang, M.~Ge, S.~Li, and J.~Dang, ``Spectrograms fusion with minimum difference masks estimation for monaural speech dereverberation,'' in \emph{Proc. ICASSP}, 2020, pp. 7544--7548.

\bibitem{shi2021spectrograms}
H.~Shi, L.~Wang, S.~Li, C.~Fan, J.~Dang, and T.~Kawahara, ``Spectrograms fusion-based end-to-end robust automatic speech recognition,'' in \emph{Proc. APSIPA ASC}, 2021, pp. 438--442.

\bibitem{luo2023music}
Y.~Luo and J.~Yu, ``Music source separation with band-split {RNN},'' \emph{{IEEE/ACM} Trans. Audio, Speech, Lang. Process.}, vol.~31, pp. 1893--1901, 2023.

\bibitem{yu2023efficient}
J.~Yu and Y.~Luo, ``Efficient monaural speech enhancement with universal sample rate band-split {RNN},'' in \emph{Proc. ICASSP}, 2023, pp. 1--5.

\bibitem{yu2023high}
J.~Yu, Y.~Luo, H.~Chen, R.~Gu, and C.~Weng, ``High fidelity speech enhancement with band-split {RNN},'' in \emph{Proc. Interspeech}, 2023, pp. 2483--2487.

\bibitem{zhang2024beyond}
W.~Zhang, K.~Saijo, J.-W. Jung, C.~Li, S.~Watanabe, and Y.~Qian, ``Beyond performance plateaus: {A} comprehensive study on scalability in speech enhancement,'' in \emph{Proc. Interspeech}, 2024, pp. 1740--1744.

\bibitem{sun2025scaling}
Z.~Sun, A.~Li, T.~Lei, R.~Chen, M.~Yu, C.~Zheng, Y.~Zhou, and D.~Yu, ``Scaling beyond denoising: {Submitted} system and findings in {URGENT} {Challenge} 2025,'' in \emph{Proc. Interspeech}, 2025, pp. 873--877.

\bibitem{luo2019conv}
Y.~Luo and N.~Mesgarani, ``{Conv-TasNet}: {Surpassing} ideal time-frequency magnitude masking for speech separation,'' \emph{{IEEE/ACM} Trans. Audio, Speech, Lang. Process.}, vol.~27, pp. 1256--1266, 2019.

\bibitem{wu2018group}
Y.~Wu and K.~He, ``Group normalization,'' in \emph{Proc. ECCV}, 2018, pp. 3--19.

\bibitem{perez2018film}
E.~Perez, F.~Strub, H.~de~Vries, V.~Dumoulin, and A.~Courville, ``{FiLM}: {Visual} reasoning with a general conditioning layer,'' in \emph{Proc. AAAI}, 2018, pp. 3942--3951.

\bibitem{dauphin2017language}
Y.~N. Dauphin, A.~Fan, M.~Auli, and D.~Grangier, ``Language modeling with gated convolutional networks,'' in \emph{Proc. ICML}, 2017, pp. 933--941.

\bibitem{dau1996quantitative}
T.~Dau, D.~P{\"u}schel, and A.~Kohlrausch, ``A quantitative model of the ``effective'' signal processing in the auditory system. {I}. {Model} structure,'' \emph{J. Acoust. Soc. Am.}, vol.~99, pp. 3615--3622, 1996.

\bibitem{dau1997modeling}
T.~Dau, B.~Kollmeier, and A.~Kohlrausch, ``Modeling auditory processing of amplitude modulation. {I}. {Detection} and masking with narrow-band carriers,'' \emph{J. Acoust. Soc. Am.}, vol. 102, pp. 2892--2905, 1997.

\bibitem{jepsen2008computational}
M.~L. Jepsen, S.~D. Ewert, and T.~Dau, ``A computational model of human auditory signal processing and perception,'' \emph{J. Acoust. Soc. Am.}, vol. 124, pp. 422--438, 2008.

\bibitem{paulick2025computational}
L.~C. Paulick, H.~Rela{\~n}o-Iborra, and T.~Dau, ``The computational auditory signal processing and perception model: {A} revised version,'' \emph{J. Acoust. Soc. Am.}, vol. 157, pp. 3232--3244, 2025.

\bibitem{lopez2001human}
E.~A. Lopez-Poveda and R.~Meddis, ``A human nonlinear cochlear filterbank,'' \emph{J. Acoust. Soc. Am.}, vol. 110, pp. 3107--3118, 2001.

\bibitem{lopez2012behavioral}
E.~A. Lopez-Poveda and P.~T. Johannesen, ``Behavioral estimates of the contribution of inner and outer hair cell dysfunction to individualized audiometric loss,'' \emph{J. Assoc. Res. Otolaryngol.}, vol.~13, pp. 485--504, 2012.

\bibitem{zaar2022predicting}
J.~Zaar and L.~H. Carney, ``Predicting speech intelligibility in hearing-impaired listeners using a physiologically inspired auditory model,'' \emph{Hear. Res.}, vol. 426, p. 108553, 2022.

\bibitem{relano2023evaluating}
H.~Rela{\~n}o-Iborra, J.~Zaar, and T.~Dau, ``Evaluating an auditory model as predictor of speech understanding in hearing-impaired listeners,'' in \emph{Proc. Forum Acust.}, 2023, pp. 861--864.

\bibitem{dubey2024icassp}
H.~Dubey, A.~Aazami, V.~Gopal, B.~Naderi, S.~Braun, R.~Cutler, A.~Ju, M.~Zohourian, M.~Tang, M.~Golestaneh, and R.~Aichner, ``{ICASSP} {2023} {Deep} {Noise} {Suppression} {Challenge},'' \emph{IEEE Open J. Signal Process.}, vol.~5, pp. 725--737, 2024.

\bibitem{panayotov2015librispeech}
V.~Panayotov, G.~Chen, D.~Povey, and S.~Khudanpur, ``{LibriSpeech}: {An} {ASR} corpus based on public domain audio books,'' in \emph{Proc. ICASSP}, 2015, pp. 5206--5210.

\bibitem{pratap2020mls}
V.~Pratap, Q.~Xu, A.~Sriram, G.~Synnaeve, and R.~Collobert, ``{MLS}: {A} large-scale multilingual dataset for speech research,'' in \emph{Proc. Interspeech}, 2020, pp. 2757--2761.

\bibitem{veaux2013voice}
C.~Veaux, J.~Yamagishi, and S.~King, ``The {Voice} {Bank} corpus: {Design}, collection and data analysis of a large regional accent speech database,'' in \emph{Proc. O-COCOSDA/CASLRE}, 2013, pp. 1--4.

\bibitem{richter2024ears}
J.~Richter, Y.-C. Wu, S.~Krenn, S.~Welker, B.~Lay, S.~Watanabe, A.~Richard, and T.~Gerkmann, ``{EARS}: {An} anechoic fullband speech dataset benchmarked for speech enhancement and dereverberation,'' in \emph{Proc. Interspeech}, 2024, pp. 4873--4877.

\bibitem{wichern2019wham}
G.~Wichern, J.~Antognini, M.~Flynn, L.~R. Zhu, E.~McQuinn, D.~Crow, E.~Manilow, and J.~Le~Roux, ``{WHAM!}: {Extending} speech separation to noisy environments,'' in \emph{Proc. Interspeech}, 2019, pp. 1368--1372.

\bibitem{fonseca2022fsd50k}
E.~Fonseca, X.~Favory, J.~Pons, F.~Font, and X.~Serra, ``{FSD50K}: {An} open dataset of human-labeled sound events,'' \emph{{IEEE/ACM} Trans. Audio, Speech, Lang. Process.}, vol.~30, pp. 829--852, 2022.

\bibitem{defferrard2017fma}
M.~Defferrard, K.~Benzi, P.~Vandergheynst, and X.~Bresson, ``{FMA}: {A} dataset for music analysis,'' in \emph{Proc. ISMIR}, 2017, pp. 316--323.

\bibitem{luo2024fast}
Y.~Luo and R.~Gu, ``Fast random approximation of multi-channel room impulse response,'' in \emph{Proc. ICASSPW}, 2024, pp. 449--454.

\bibitem{sondergaard2013auditory}
P.~S{\o}ndergaard and P.~Majdak, ``The auditory modeling toolbox,'' in \emph{The Technology of Binaural Listening}.\hskip 1em plus 0.5em minus 0.4em\relax Springer, 2013, pp. 33--56.

\bibitem{majdak2022amt}
P.~Majdak, C.~Hollomey, and R.~Baumgartner, ``{AMT} 1.x: {A} toolbox for reproducible research in auditory modeling,'' \emph{Acta Acust.}, vol.~6, p.~19, 2022.

\bibitem{bradley2003importance}
J.~S. Bradley, H.~Sato, and M.~Picard, ``On the importance of early reflections for speech in rooms,'' \emph{J. Acoust. Soc. Am.}, vol. 113, pp. 3233--3244, 2003.

\bibitem{roman2013speech}
N.~Roman and J.~Woodruff, ``Speech intelligibility in reverberation with ideal binary masking: {Effects} of early reflections and signal-to-noise ratio threshold,'' \emph{J. Acoust. Soc. Am.}, vol. 133, pp. 1707--1717, 2013.

\bibitem{zeghidour2021wavesplit}
N.~Zeghidour and D.~Grangier, ``Wavesplit: {End}-to-end speech separation by speaker clustering,'' \emph{{IEEE/ACM} Trans. Audio, Speech, Lang. Process.}, vol.~29, pp. 2840--2849, 2021.

\bibitem{choi2022empirical}
S.~Choi, Y.~Lee, J.~Park, H.~Y. Kim, B.-Y. Kim, Z.-Q. Wang, and S.~Watanabe, ``An empirical study of training mixture generation strategies on speech separation: {Dynamic} mixing and augmentation,'' in \emph{Proc. APSIPA ASC}, 2022, pp. 1070--1075.

\bibitem{graetzer2022dataset}
S.~Graetzer, M.~A. Akeroyd, J.~Barker, T.~J. Cox, J.~F. Culling, G.~Naylor, E.~Porter, and R.~Viveros~Mu{\~n}oz, ``Dataset of {British} {English} speech recordings for psychoacoustics and speech processing research: {The} {Clarity} speech corpus,'' \emph{Data Br.}, vol.~41, p. 107951, 2022.

\bibitem{mesaros2016tut}
A.~Mesaros, T.~Heittola, and T.~Virtanen, ``{TUT} database for acoustic scene classification and sound event detection,'' in \emph{Proc. EUSIPCO}, 2016, pp. 1128--1132.

\bibitem{kingma2015adam}
D.~P. Kingma and J.~Ba, ``Adam: {A} method for stochastic optimization,'' in \emph{Proc. ICLR}, 2015.

\bibitem{bisgaard2010standard}
N.~Bisgaard, M.~S. M.~G. Vlaming, and M.~Dahlquist, ``Standard audiograms for the {IEC} 60118-15 measurement procedure,'' \emph{Trends Amplif.}, vol.~14, pp. 113--120, 2010.

\bibitem{rix2001perceptual}
A.~W. Rix, J.~G. Beerends, M.~P. Hollier, and A.~P. Hekstra, ``Perceptual evaluation of speech quality ({PESQ})--{A} new method for speech quality assessment of telephone networks and codecs,'' in \emph{Proc. ICASSP}, 2001, pp. 749--752.

\bibitem{jensen2016algorithm}
J.~Jensen and C.~H. Taal, ``An algorithm for predicting the intelligibility of speech masked by modulated noise maskers,'' \emph{{IEEE/ACM} Trans. Audio, Speech, Lang. Process.}, vol.~24, pp. 2009--2022, 2016.

\bibitem{kates2014hearing}
J.~M. Kates and K.~H. Arehart, ``The hearing-aid speech perception index ({HASPI}),'' \emph{Speech Commun.}, vol.~65, pp. 75--93, 2014.

\bibitem{kates2010hearing}
------, ``The hearing-aid speech quality index ({HASQI}),'' \emph{J. Audio Eng. Soc.}, vol.~58, pp. 363--381, 2010.

\bibitem{barker2023pyclarity}
\BIBentryALTinterwordspacing
J.~Barker, G.~Roa-Dabike, Z.~Tu, and N.~Shephard, ``pyclarity,'' The University of Sheffield, 2023. [Online]. Available: \url{https://github.com/claritychallenge/clarity}
\BIBentrySTDinterwordspacing

\bibitem{byrne1986national}
D.~Byrne and H.~Dillon, ``The {National} {Acoustic} {Laboratories}' ({NAL}) new procedure for selecting the gain and frequency response of a hearing aid,'' \emph{Ear Hear.}, vol.~7, pp. 257--265, 1986.

\bibitem{leroux2019sdr}
J.~Le~Roux, S.~Wisdom, H.~Erdogan, and J.~R. Hershey, ``{SDR} -- {Half}-baked or well done?'' in \emph{Proc. ICASSP}, 2019, pp. 626--630.

\bibitem{kinoshita2020improving}
K.~Kinoshita, T.~Ochiai, M.~Delcroix, and T.~Nakatani, ``Improving noise robust automatic speech recognition with single-channel time-domain enhancement network,'' in \emph{Proc. ICASSP}, 2020, pp. 7009--7013.

\bibitem{welch1967use}
P.~D. Welch, ``The use of fast {Fourier} transform for the estimation of power spectra: {A} method based on time averaging over short, modified periodograms,'' \emph{{IEEE} Trans. Audio Electroacoust.}, vol.~15, pp. 70--73, 1967.

\bibitem{jepsen2011characterizing}
M.~L. Jepsen and T.~Dau, ``Characterizing auditory processing and perception in individual listeners with sensorineural hearing loss,'' \emph{J. Acoust. Soc. Am.}, vol. 129, pp. 262--281, 2011.

\bibitem{relano2022speech}
H.~Rela{\~n}o-Iborra and T.~Dau, ``Speech intelligibility prediction based on modulation frequency-selective processing,'' \emph{Hear. Res.}, vol. 426, p. 108610, 2022.

\bibitem{wang2005ideal}
D.~Wang, ``On ideal binary mask as the computational goal of auditory scene analysis,'' in \emph{Speech Separation by Humans and Machines}.\hskip 1em plus 0.5em minus 0.4em\relax Springer, 2005, pp. 181--197.

\bibitem{wang2006computational}
D.~Wang and G.~J. Brown, \emph{Computational auditory scene analysis: Principles, algorithms, and applications}.\hskip 1em plus 0.5em minus 0.4em\relax Wiley-IEEE Press, 2006.

\bibitem{hassager2017preserving}
H.~G. Hassager, T.~May, A.~Wiinberg, and T.~Dau, ``Preserving spatial perception in rooms using direct-sound driven dynamic range compression,'' \emph{J. Acoust. Soc. Am.}, vol. 141, pp. 4556--4566, 2017.

\bibitem{may2018signal}
T.~May, B.~Kowalewski, and T.~Dau, ``Signal-to-noise-ratio-aware dynamic range compression in hearing aids,'' \emph{Trends Hear.}, vol.~22, p. 2331216518790903, 2018.

\bibitem{may2020scene}
------, ``Scene-aware dynamic-range compression in hearing aids,'' in \emph{The Technology of Binaural Understanding}.\hskip 1em plus 0.5em minus 0.4em\relax Springer, 2020, pp. 763--799.

\bibitem{raichle2026test}
T.~Raichle, E.~Amini, and B.~Yang, ``Test-time adaptation for speech enhancement via mask polarization,'' in \emph{Proc. ICASSP}, 2026, pp. 18\,882--18\,886.

\bibitem{wang2014training}
Y.~Wang, A.~Narayanan, and D.~Wang, ``On training targets for supervised speech separation,'' \emph{{IEEE/ACM} Trans. Audio, Speech, Lang. Process.}, vol.~22, pp. 1849--1858, 2014.

\end{thebibliography}

\end{document}